\documentclass[article,twocolumn,showpacs,superscriptaddress,10pt]{revtex4-1} 

\usepackage{amsmath,amssymb,graphicx}
\usepackage{mathbbol}
\usepackage[dvipsname]{xcolor}
\usepackage{upgreek}

\newcommand{\TextTh}{\textsuperscript{th}}

\newcommand{\micron}{\ensuremath{\upmu\text{m}}}
\newcommand{\logo}{{\sf PAINTER}} 		
\newcommand{\ii}{\text{i}} 				
\newcommand{\indic}{\mathbb{1}} 			
\newcommand{\Id}{\it{\bf{I}}} 			
\newcommand{\un}{\it{\bf{1}}} 				
\newcommand{\tr}{\text{tr}} 		
\newcommand{\bB}{{\bf{b}}} 			
\newcommand{\ux}{\textbf{x}} 				
\newcommand{\uX}{\textbf{X}} 				
\newcommand{\uy}{\textbf{y}} 				
\newcommand{\uyt}{\widetilde{\textbf{y}}} 		
\newcommand{\uY}{{{\mathrm{\textbf{Y}}}}} 	
\newcommand{\bH}{\textbf{H}} 				
\newcommand{\bh}{\textbf{h}} 				
\newcommand{\uF}{\textbf{F}} 				
\newcommand{\uuF}{{\uF}} 				
\newcommand{\uphi}{\boldsymbol{\varphi}} 		
\newcommand{\upsi}{\boldsymbol{\psi}} 		
\newcommand{\uDphi}{{\boldsymbol\Delta  \boldsymbol{\varphi}}}  		
\newcommand{\bpsi}{\boldsymbol{\psi}} 		
\newcommand{\bdphi}{{\Delta \, \varphi}}  	
\newcommand{\uxi}{\boldsymbol{\xi}} 			
\newcommand{\upow}{\boldsymbol{\gamma}} 	
\newcommand{\uPow}{\boldsymbol{\Gamma}} 
\newcommand{\datapow}{\boldsymbol{\zeta}} 	
\newcommand{\unoise}{\boldsymbol{\eta}} 	
\newcommand{\uw}{\boldsymbol{\omega}} 	
\newcommand{\TotV}{\text{spat}} 			
\newcommand{\SpecS}{\text{spec}} 			
\newcommand{\Reg}{\boldsymbol{\Omega}} 	
\newcommand{\bM}{\textbf{M}} 			
\newcommand{\tP}{P} 			
\newcommand{\tT}{T} 			
\newcommand{\tS}{S} 			
\newcommand{\tV}{V} 			
\newcommand{\tU}{U} 			
\newcommand{\tD}{D} 			
\newcommand{\uD}{\textbf{\tD}} 			
\newcommand{\uP}{\textbf{\tP}} 	
\newcommand{\uT}{\textbf{\tT}} 	
\newcommand{\uV}{\textbf{\tV}} 	
\newcommand{\uS}{\textbf{\tS}} 	
\newcommand{\uU}{\textbf{\tU}} 	
\newcommand{\bHtv}{\textbf{H}^\TotV } 		
\newcommand{\mutv}{\mu_s} 		
\newcommand{\ptv}{{\text{p}_\TotV}} 		
\newcommand{\bHss}{\textbf{H}^\SpecS} 		
\newcommand{\muss}{\mu_\lambda} 	
\newcommand{\pss}{{\text{p}_\SpecS}} 		
\newcommand{\mutvss}{\mu_{s \slash \lambda}} 		
\newcommand{\ptvss}{{\text{p}_{\TotV \slash \SpecS}}} 	
\newcommand{\uC}{\textbf{C}} 	
\newcommand{\support}{\boldsymbol{\Pi}} 
\newcommand{\utau}{\boldsymbol{\upsilon}} 	
\newcommand{\uTau}{\boldsymbol{\Upsilon}} 	
\newcommand{\beq}{\begin{eqnarray}}
\newcommand{\eeq}{\end{eqnarray}}
\newcommand{\beqs}{\begin{eqnarray*}}
\newcommand{\eeqs}{\end{eqnarray*}}
\newcommand{\lb}{\left [}
\newcommand{\rb}{\right ]}
\newcommand{\lbr}{\left (}
\newcommand{\rbr}{\right )}

\begin{document}

\title{PAINTER: a spatio-spectral image reconstruction\\ algorithm for optical interferometry}

\author{Antony Schutz}\email{Corresponding author: aschutz@oca.eu} 

\author{Andr\'e Ferrari}
\author{David Mary}
\affiliation{Lab. J.-L. Lagrange, Universit\'e de Nice Sophia Antipolis, CNRS, Observatoire de la C\^ote d'Azur, Parc Valrose, F-06108 Nice cedex 02, France}

\author{Ferr\'eol Soulez}
\author{\'Eric Thi\'ebaut}
\affiliation{Universit\'e de Lyon, Lyon F-69003, France. Universit\'e Lyon 1, Observatoire de Lyon, 9 avenue Charles Andr\'e, Saint-Genis Laval F-69230, France. CNRS, UMR 5574, Centre de Recherche Astrophysique de Lyon, \'Ecole Normale Sup\'erieure de Lyon, Lyon F-69007, France}

\author{Martin Vannier}
\affiliation{Lab. J.-L. Lagrange, Universit\'e de Nice Sophia Antipolis, CNRS, Observatoire de la C\^ote d'Azur, Parc Valrose, F-06108 Nice cedex 02, France}


\begin{abstract}
Astronomical optical interferometers  sample the  Fourier transform of the intensity distribution of a source
 at the observation wavelength.
Because of rapid perturbations caused by atmospheric turbulence,
the phases of the complex Fourier samples (visibilities) cannot be directly exploited. 
Consequently, specific image reconstruction methods have been devised in the last few decades. 
Modern polychromatic optical interferometric instruments are now paving the way to multiwavelength imaging. 
This paper is devoted to the derivation of a spatio-spectral  (``3D'')  image reconstruction algorithm, coined 
\logo \ (Polychromatic opticAl INTErferometric Reconstruction software). 
The algorithm relies on an iterative process, which alternates  estimation of polychromatic images and of  complex visibilities.
The complex visibilities are not only estimated
 from squared  moduli and closure phases, but also differential phases, which helps to better constrain the polychromatic reconstruction.
Simulations on synthetic data illustrate the efficiency of the algorithm and in particular the relevance 
of injecting a differential phases model in the reconstruction.
\end{abstract}


\maketitle 

\section{Introduction} \label{sec:Introduction}

Current astronomical Optical Interferometers (OI) do not directly provide images, even if prospective studies exist in this direction \cite{lab1996,aime2012,2012SPIE2}. Instead, OI data  are related to the observed celestial scene at the observation wavelength by a two dimensional Fourier transform (FT) of the corresponding intensity distribution perpendicular to the line of sight. 
Ideally, the observables measured by an interferometer are the so-called \textit{complex visibilities}, which corresponds to complex samples of the Fourier spectrum.
The  sampling function is fully defined by the positions of the interfering telescopes and by the observation wavelength.  Earth rotation provides additional samples for observations acquired at different epochs, as the telescopes' positions are modified in time with respect to the line of sight (an effect called super synthesis). However, owing to the small number of telescopes  involved (typically three or four), the sampling of the Fourier space that is achieved by OI is  always sparse.

In essence, measuring moduli and phases of complex visibilities in OI amounts to measuring contrasts and phases of interference fringes \citep{labeyrie75}. 
Atmospheric turbulence randomly shifts  these fringes on a timescale of $10$ ms.
The moduli are thus obtained by estimating fringe contrasts in snapshot mode with short integration times 
that freeze the atmospheric turbulence. As for the phases, the unknown random phase shifts imply that
 optical interferometers cannot measure directly the phases
 (in absence of an artificial or real star, which would provide such a reference \cite{monnier2003}).
Astronomers thus use a turbulence independent observable related to the phases called  \textit{closure phase},
originally devised for radiointerferometers \cite{jennison58}, and
from which the phase information can   be  partially extracted 
\cite{VisEst1984,CP1986,VisEst1994}. Note that the current situation in optical is however very different from radio,
 where the numbers of antennas or dipoles is several orders of magnitude greater than in optical (see e.g. \cite{MPvanHaarlem:2013gi} for a recent example) and all phases can be  estimated.

From an informational viewpoint, the image reconstruction problem posed by OI is highly underdetermined because of the sparse Fourier sampling and of additional missing phase information. This means that formally, 
infinitely many intensity distributions exist that are consistent with OI data. In this framework, a classical and well-understood
strategy for image reconstruction is to adopt  an inverse problem approach, where missing information is mitigated, and hopefully compensated for, by \textit{a priori} knowledge \cite{thiebauteas}. In this case, the image reconstruction algorithm aims at finding an
intensity distribution that minimizes a cost function composed of a data fidelity term, which is related to the noise distribution, plus a regularization term and possibly other constraints, which are related to  prior knowledge.

Following this path, various algorithms have blossomed in the last twenty years.
Most of the proposed algorithms rely on gradient descent methods (\textit{WISARD} \cite{meimon2005,Meimon:05}, \textit{BSMEM} \cite{BSMEM}, \textit{MiRA} \cite{MIRA},  \textit{BBM} \cite{BBM}, \textit{IRBis} \cite{IRBis}).
A different approach is used in \textit{MACIM} \cite{MACIM} and in its evolution \textit{SQUEEZE} \cite{baron2012},
which rely on Markov Chain Monte Carlo (MCMC) method.

While these algorithms have proved very useful to astronomers, none of them is currently able to tackle 
polychromatic reconstruction for general sources having variations in their intensity distribution in wavelength. As such, they are monochromatic imagers and are not applicable for multiwavelength image reconstruction. 

Nowadays,  modern  OI is  however polychromatic (see for instance  AMBER \cite{AMBER2000SPIE}, PIONIER \cite{PIONIER2011}, or VEGA \cite{Vega2009}) and more powerful polychromatic instruments are in development (like MATISSE \cite{MATISSE} and GRAVITY \cite{GRAVITY}). In such devices, interference fringes are simultaneously recorded in a number of wavelength channels that can reach several hundreds. Indeed, the fundamental justification of polychromatic observations is that the amount and distribution of electromagnetic radiation emitted by astrophysical sources may be very different from on wavelength channel to another,
as  light-matter interactions are highly variable in wavelength (emission and absorption lines for instance). Multiwavelength
 OI thus constitutes an extremely rich evolution over monochromatic OI. In order to fully exploit these instruments, the multiwavelength evolution of the mature but monochromatic OI reconstruction algorithms is mandatory.

Building on the monochromatic approaches mentioned above, some first steps have recently been undertaken in the direction of multiwavelength reconstruction. The work \cite{sparco} implements a
semi-parametric algorithm for the image reconstruction of chromatic objects,  dedicated to the case of
 a central object surrounded by an  extended structure such as a young star. The approach of \cite{admmthiebaut,admmsoulez} 
 uses a sparsity-regularized approach, dedicated to the case  where the observed scene is a collection of point-like sources.
\cite{SelfCal} uses MiRA as a key optimization engine combined to a Self Calibration approach
and demonstrates the potential of using the differential phases in an image-reconstruction process.

This paper is devoted to the derivation of a  multiwavelength or spatio--spectral images reconstruction (3D) algorithm named \logo \ (for Polychromatic opticAl INTErferometric Reconstruction software). 
This approach uses the absolute visibilities and closure phases, which are
 considered dependent of the wavelength. In addition, we also use the so-called \textit{differential phases}, which
 are defined as the  phases relatively to a reference channel and constitute an additional turbulence independent observable of the phases  in multiwavelength observation mode.

The algorithm relies on the alternate estimation of the complex visibilities (from estimated phases and observed noisy moduli)
and of the polychromatic intensity distribution (using spatio-spectral regularization and constraints).
From a modeling viewpoint, the main originality of the approach is to estimate the unknown phases from both closure phases and differential phases.
From an optimization viewpoint, the  algorithm  is based on ADMM methodology \cite{ADMMBoyd}.
 \logo  \ can be seen as an evolution of the \textit{MiRA--3D} algorithm proposed in \cite{admmthiebaut,admmsoulez}.
An implementation of  \logo  \  in matlab (octave compatible) with input data in OI-FITS format \cite{OIFITS},  
is available at \url{https://www-n.oca.eu/aferrari/painter}.

The paper is organized as follows: 
sections \ref{sec:notations} and \ref{sec:DataModeling} introduce notations and data modeling. We derive here
 an extended model of phase differences which is specific to the 3D reconstruction.
Section \ref{sec:InversePb}  tackles the inverse problem approach. We
introduce  assumptions related to the data fidelity criterion (noise perturbations) as well as prior knowledge in the form of regularizations and constraints. 
Section \ref{sec:3Dimages} derives the resulting 3D image reconstruction algorithm.
Performances of the algorithm are provided and analyzed  in section \ref{sec:simus}.

\section{Notations} \label{sec:notations}
\begin{center}
\begin{tabular}{r l}
$\angle\; \cdot$ 	\;\;&\;\;  phases of  complex numbers \\ \;\;&\;\;  (may result in a scalar, a vector or a matrix) \\
$|\cdot|$ 	\;\;&\;\;  moduli of  complex numbers \\ \;\;&\;\;  (may result in a scalar, a vector or a matrix) \\
$^\ast$ \;\;&\;\; complex conjugate 					\\
$^\top$ \;\;&\;\; transpose \\
$\widehat{\;}$ \;\;&\;\; Fourier transform						\\ 
$^H$ \;\;&\;\; complex conjugate transpose  			\\
$\uX$ \;\;&\;\; matrix   \\ 
$\ux$ 	\;\;&\;\; vector, with components 	$\ux_n$	\\
 $\otimes$ \;\;&\;\;  Kronecker product (tensor product) 		\\
$\oplus_n^N$ \;\;&\;\;  direct sum: $\oplus_n^N \uX_n = \text{block diag} \lbr \uX_1, \ldots, \uX_N \rbr$ \\\;\;&\;\; i.e.  $\oplus_{n=1}^2 \uX_n = \uX_1 \oplus \uX_2 = \lb 
  \begin{array}{ c c }
     \uX_1 & 0 \\
     0 & \uX_2
  \end{array}  \rb$ \\
$\odot$ \;\;&\;\;  Hadamard product (dot product) 		\\
$\Id_N$ \;\;&\;\;  identity matrix of size $N \times N$ 		\\
$\un_N$ \;\;&\;\; vector $[1,\hdots,1]^\top$  of length $N$ 		\\
$\indic_{\mathbb{R}^+}$  \;\;&\;\;  indicator function on positive orthant 		\\
$\tr \lbr \uX \rbr$ \;\;&\;\;   trace of $\uX$ \\
$ \text{vec}\, \uX$ \;\;&\;\; matrix vectorization i.e the columns of $\uX$ \\ \;\;&\;\; are stacked into one column vector $\ux$\\
$\text{diag}(\ux)$ \;\;&\;\; diagonal matrix with $\ux$ on its diagonal \\
$\| \uX \|_\text{F}^2$ \;\;&\;\;  squared Frobenius norm \\
$\| \ux \|_2$ \;\;&\;\;  $\ell_2$ norm \\
$\ux^+$   \;\;&\;\; update of $\ux$
\end{tabular}
\end{center}

\section{Data modeling} \label{sec:DataModeling}
\subsection{Spatio-spectral model}
In the absence of atmospheric turbulence the observable measured by an interferometer is the complex visibility \cite{thiebauteas}.
This observable is measured from the fringe pattern obtained by the interference of two beams collected from a pair of telescopes. The spatial position of each such pair defines one of the $N_\bB$ baselines of the telescope array. 
Hereafter, the notation for the baseline $\bB_{a,b}$ refers to the position vector of a telescope pair $(a,b)$ projected on a plane
perpendicular to the line of sight.

In the considered case of polychromatic observations,
an astrophysical source is described by  an intensity distribution which is a function of wavelength.
Because OI instrument always have limited fields of view, we assume that  the distribution of interest accounts for
the limited spatial response of the interferometer: it is an apodized version of the intensity distribution around the line of sight.
The unkown distribution can be written as
 $ I(\delta,\tau,\lambda)=I_\lambda(\delta,\tau)$, which is a flux density at angular position $\boldsymbol{\theta}=[\delta,\tau]^\top$ of the sky
and wavelength $\lambda$. 

 In absence of any perturbation and for purely monochromatic observations, a telescope pair  of baseline $\bB$ provides a complex visibility defined by   $y^\lambda=\widehat{I}(\frac{\bB}{\lambda})$. 
Because OI instruments have limited angular and spectral resolutions (respectively set by the maximum distance between two telescopes and by the bandpass of the optical filters), a simple way to represent the unknown spatio-spectral distribution of the sources
is to discretize $ I(\delta,\tau,\lambda)$ over \textit{voxels}. We consider here for simplicity the same discretization $ \Delta \delta = \Delta \tau =\Delta \theta$ in both angular coordinates, resulting in $N_x \times N_x$ parameters per image, and an instrument
 with $N_\lambda$ wavelength channels  of equal bandwidth $\Delta \lambda$, which is set to the spectral resolution.
In this case, all voxels have the same size  $\Delta \theta \times  \Delta \theta \times \Delta \lambda$. If we further assume a unit transfer function in all channels, one voxel is simply defined as:  
\beq \ux_i^{\lambda_n}= \int_{\delta_l}^{\delta_l+\Delta \theta}\int_{\tau_m}^{\tau_m+\Delta \theta}\int_{\lambda_n}^{\lambda_n+\Delta \lambda} I(\delta,\tau,\lambda)\; \text{d}\boldsymbol{\theta}  \text{d}\lambda
\eeq
where $i$ refers to a pixel at angular position $[\delta_l,\tau_m]$ and $\lambda_n$ to the reference wavelength of channel $n$.
The column vector $\ux^{\lambda_n}$ collects all voxels and can be organized as the vectorization of a $\lbr N_x \times N_y \rbr$ image of the astrophysical  source at wavelength $\lambda_n$.
In this setting, the goal of the multiwavelength reconstruction algorithm is  to estimate the voxels, which represent the unknown parameters of the model.
 
Let  $\text{y}^{\lambda_n}_{a,b} $ be the complex visibility  at the spatial frequency  ${\bB}\slash{\lambda_n}$
 , and let  $ \uy^{\lambda_n}$ be the column vector collecting the set of complex visibilities corresponding to all available baselines at wavelength $\lambda_n$. 
The complex visibilities can then be related in matrix form to the parameters by the direct model \cite{Thiebaut2010,admmthiebaut}
\beq
\uy^{\lambda_n} = \uF^{\lambda_n} ~\ux^{\lambda_n}  \label{eq:yfx}
\eeq
where $\uF^{\lambda_n}$ is obtained from a Non Uniform Discrete Fourier Transform (NuDFT) \cite{fessler2003} at the spatial frequencies imposed by the geometry of the telescope array and by the observation wavelength $\lambda_n$.

The previous expression describes the complex visibilities by wavelength. A  compact notation including all wavelengths and baselines is
\beq
\uy  &=&  \uuF ~\ux,\; \uuF = \oplus^{N_\lambda}_{n=1} \, \uF^{\lambda_n}  \label{eq:TensorEq} \\
\ux &=& \lb {\ux^{\lambda_1}}^\top, \ldots, {\ux^{\lambda_{N_\lambda}}}^\top \rb^\top \nonumber 
\eeq
where $\uuF $ is a block diagonal matrix with each block referring to the NuDFT at a particular wavelength.
Vector  $\uy$ concatenates the complex visibility vectors ($\uy^{\lambda_n} $ of Eq.~\ref{eq:yfx}) for all wavelengths into a $ N_\bB N_\lambda \times 1 $ visibility vector, with associated moduli  $\upow$ and phases $\uphi$ given by
\beq 
\uy &=& \lb {\uy^{\lambda_1}}^\top, \ldots, {\uy^{\lambda_{N_\lambda}}}^\top \rb^\top \label{eq:vis}\\
\upow &=& \lb {\upow^{\lambda_1}}^\top, \ldots, {\upow^{\lambda_{N_\lambda}}}^\top \rb^\top = | \uy  | \label{eq:pow} \\
\uphi &=& \lb {\uphi^{\lambda_1}}^\top, \ldots, {\uphi^{\lambda_{N_\lambda}}}^\top \rb^\top = \angle\;~\uy \label{eq:phi} 
\eeq
In order to analyze the chromatic variation of the visibilities $\uy^{\lambda_n}_m$ and of the images ${\ux^{\lambda_n}}$ over the $N_\lambda$ wavelengths, 
we also need to introduce the  $ N_\bB \times N_\lambda   $ matrix $\uY$ and the $ N^2_x \times N_\lambda $
matrix $\uX$ defined by
\beq 
\uY &=& \lb {\uy^{\lambda_1}}, \ldots, {\uy^{\lambda_{N_\lambda}}} \rb=\; \textrm{vec}^{-1} \;\uy \label{eq:ModMatY}\\
\uX &=& \lb {\ux^{\lambda_1}}, \ldots, {\ux^{\lambda_{N_\lambda}}} \rb\; = \textrm{vec}^{-1} \;\ux \, . \label{eq:ModMatX}
\eeq
To clarify the use of a matrix notation note that the $n\TextTh$ column of $\uX$ corresponds to the vectorization of the image at the wavelength $\lambda_n$ while the $p\TextTh$ line is for the variation of the pixel $p$ along the wavelengths.

\subsection{Model of phase differences} \label{sec:ModelPhaseDiff}

In the presence of atmospheric turbulence, the beams received at each telescope  are affected by random and different optical path differences, which corrupt the phases measurements of the complex visibilities.
To overcome this difficulty, turbulence independent quantities need to be constructed. 

\subsubsection{ Closure phase} \label{sec:T3Phi}
The first phase difference information used for image reconstruction in presence of turbulent measurements is the closure phase (operator $\psi$). 
It is defined as the phase of the bispectrum \cite{CP1986}, i.e., the Fourier transform of the triple correlation.
 For three baselines $\lb \bB_{a,b}, \bB_{b,c}, \bB_{a,c} \rb$ corresponding to a triplet $(a,b,c)$ of telescopes, the ``atmospheric corrupted'' {\it instantaneous} visibilities at a given wavelength $\lambda_n$  can be modeled as
\beq
 \text{y}^{\lambda_n}_{a,b} &=& \gamma^{\lambda_n}_{a,b} \, \exp\lbr\ii\,[  \varphi^{\lambda_n}_{a,b} + \eta^{\lambda_n}_{a} - \eta^{\lambda_n}_{b}   ]\rbr \label{eq:TurbAtmo1}\\
 \text{y}^{\lambda_n}_{b,c} &=& \gamma^{\lambda_n}_{b,c} \, \exp\lbr\ii\,[ \varphi^{\lambda_n}_{b,c} + \eta^{\lambda_n}_{b} - \eta^{\lambda_n}_{c}   ]\rbr  \label{eq:TurbAtmo2} \\
 \text{y}^{\lambda_n}_{a,c} &=& \gamma^{\lambda_n}_{a,c} \, \exp\lbr\ii\,[ \varphi^{\lambda_n}_{a,c} + \eta^{\lambda_n}_{a} - \eta^{\lambda_n}_{c}   ]\rbr  \label{eq:TurbAtmo3}
\eeq
where $\varphi^{\lambda_n}_\cdot$ are the quantities of interest (i.e., the uncorrupted  phases) and $\eta^{\lambda_n}_\cdot$ are perturbation terms related to the corresponding telescopes. The  closure phase associated with this triplet is defined as
\beq
\psi^{\lambda_n}_{a,b,c}  &=& \angle\; \text{y}^{\lambda_n} _{a,b } ~\text{y}^{\lambda_n} _{b,c } ~{\text{y}^{\lambda_n}_{a,c}}^\ast  \nonumber \\
&=& \varphi^{\lambda_n}_{a,b} \,+\, \varphi^{\lambda_n}_{b,c}  \,-\, \varphi^{\lambda_n}_{a,c} \label{eq:T3Diff} \\
 &=&  \bh^{\lambda_n}_{a,b,c}  \uphi^{\lambda_n}  \label{eq:hphi}
\eeq
 where $\uphi^{\lambda_n}$  is as in Eq. \ref{eq:phi}
the vector containing all \textit{unperturbed} phases for  wavelength $\lambda_n$, and 
 $\bh^{\lambda_n}_{a,b,c}$ is a sparse row vector with only three non zeros entries which take values $\{1, 1, -1\}$ (as reflected by Eq.~\ref{eq:T3Diff}). Clearly, the phase closure allows to get rid of atmospheric effects for triplets of complex visibilities. If $N_t$ denotes the number of telescopes there are ${(N_t-1)(N_t-2)}\slash{2}$  independent closure phases \cite{AdvancedImaging}.

Let matrix $\bH^{\lambda_n}$  concatenate in its rows the independent closure phases of the type (\ref{eq:hphi}) that can be obtained for  the available triplets of telescopes  at wavelength $\lambda_n$. 
For the simplicity of the presentation but without loss of generality, we assume here that all telescope pairs observe the same channels. In this case,  the   triplets  involved in independent closure phases can be taken as the same in each wavelength channel. The global closure phase operator
 $\bH^{\bpsi}$ is then simply a block diagonal matrix
that replicates $ \bH^{\lambda_1}$, the closure phase matrix for $\lambda_1$
\begin{equation}
\upsi = \bH^{\bpsi} ~ \uphi ,\; \bH^{\bpsi} = \Id_{N_\lambda} \otimes \bH^{\lambda_\textrm{1}} \label{eq:clotphases}
\end{equation}
where $\upsi$ is the vector of all closure phases and  $\uphi $  is the unknown unperturbed phase vector of Eq. \ref{eq:phi}.
\subsubsection{Differential Phases} \label{sec:DiffPhi}
A second phase difference information that is particularly interesting in polychromatic imaging is the differential phase $\Delta \varphi$.
 For  one baseline $\bB_{a,b}$, differential phases measure the phase evolution in wavelength   with respect to a phase reference channel   (see e.g. \cite{AMBER}). They are defined as
 \begin{eqnarray}
 \bdphi^{\lambda_k, \lambda_\text{ref}}_{{a,b}} & =& \angle\; \text{y}^{\lambda_k} _{a,b }-\angle\; \text{y}^{\lambda_{\textrm{ref}}} _{a,b } \\
 &=& \varphi^{\lambda_k}_{a,b} +  \eta^{\lambda_k}_{a} - \eta^{\lambda_k}_{b}   - \left(\varphi^{\lambda_\text{ref}}_{a,b} +  \eta^{\lambda_\text{ref}}_{a} - \eta^{\lambda_\text{ref}}_{b} \right) \nonumber
\label{eq:phaseDiff0}
\end{eqnarray}
 If the analyzed bandwidth is relatively narrow, which is generally the case, the phase turbulence terms 
 on each telescope  $\eta^{\lambda_k}_{\cdot}$ and $\eta^{\lambda_\text{ref}}_{\cdot}$
are in first approximation independent of the wavelength \cite{DiffVis2012}. Thus the phase difference becomes
\beq
 \bdphi^{\lambda_k, \lambda_\text{ref}}_{{a,b}}  = \varphi^{\lambda_k}_{a,b}  - \varphi^{\lambda_\text{ref}}_{a,b}  \label{eq:phaseDiff}
\eeq
As closure phases,  the differential phases in Eq.~\ref{eq:phaseDiff} are essentially not  affected by the atmospheric perturbation.  

The reference channel can be chosen as one of the available wavelengths. 
In this case $N_\lambda -1$ differential phases are available per baseline.
Without loss of generality, we denote below by $\lambda_1$ the reference channel.
Similarly to Eq.~\ref{eq:hphi} we can write using definition of  $\uphi$ in Eq.~\ref{eq:phi}: 
\begin{equation}
\bdphi^{\lambda_k, \lambda_1}_{{a,b}}  =  \bh^{\lambda_k, \lambda_1}_{a,b}  \uphi \label{eq:hdphi}
\end{equation}
where $\bh^{\lambda_k, \lambda_1}_{a,b}$ is a sparse row vector with only two non zeros entries which take values $\{1, -1\}$.
Denoting by  $\uDphi$ the differential phases vector which collects all wavelengths differences for all baselines, and by
$\bH^{\uDphi}$ the matrix which concatenates all vectors $  \bh^{\lambda_k, \lambda_1}_{a,b}$ in its rows,
we have similarly to Eq. \ref{eq:clotphases}
\begin{equation}
\uDphi = \bH^{\uDphi} ~ \uphi, \;  \bH^{\uDphi} = \lbr - \un_{(N_\lambda -1)} \otimes \left.  \Id_{N_\bB}  
\right|   \Id_{(N_\lambda -1)\times N_\bB} \rbr \label{eq:diffphase}
\end{equation}
Equations \ref{eq:clotphases} and \ref{eq:diffphase} can be merged in the single measurement
equation
\begin{equation}
\uxi = \bH ~\uphi \label{eq:phasecomp}
\end{equation}
where $\bH=\lb \frac{\bH^{\bpsi}}{\bH^{\uDphi} } \rb$ models all phase difference information.

\subsection{Practical remarks}
In practice, a specific phase reference channel $\varphi^{\lambda_\text{ref}}_{a,b}$ 
is sometimes computed using all the visibilities associated with a  baseline. One possibility is
to set the phase reference as the angle of the empirical mean over the channels:  
$$ \varphi^{\lambda_\text{ref}}_{a,b} = \angle\; 
\frac{1}{ N_\lambda} \sum^{N_\lambda}_{n=1} \text{y}^{\lambda_n} _{a,b}$$
Another possibility, as for the AMBER instruments \cite{DiffVis2012,millour2006} is to define a channel dependent reference: 
$$ \varphi^{\lambda_\text{ref}}_{a,b} = \angle\; 
\frac{1}{ N_\lambda-1} \sum^{N_\lambda}_{m\neq n} \text{y}^{\lambda_m} _{a,b}$$
In these cases, the formalism of Eq.~\ref{eq:phaseDiff} is still valid assuming that all the reference 
channels for a given baseline are almost equal and will cancel in the difference of differential phases.
Equation \ref{eq:phaseDiff} can then be replaced by:
\beq
\bdphi^{\lambda_2, \lambda_1}_{a,b}  =  \bdphi^{\lambda_2,\lambda_\text{ref}}_{a,b} 
- \bdphi^{\lambda_1,\lambda_\text{ref}}_{a,b}  \label{eq:phaseDiffRef}
\eeq
and the linear Eq.~\ref{eq:diffphase} is kept  with the same matrix $\bH^{\uDphi}$.

Consider the general case where $N_t > 3$ telescopes configurations are used to observe at $N_\lambda>1$ wavelengths.
For a single exposure the number of spatial visibilities measured and so the number of phases to estimate is $N_\phi=N_\bB N_\lambda$, where the number of bases $N_\bB  = N_t(N_t-1) \slash 2 $. 
If the matrix $\bH$ involved in the phases to the phases differences transformation is full ranked then there are no phases identifiability ambiguities.
In practice, this matrix is the concatenation of two matrices, the first for the phase closures and the second for the differential phases.

A closure is said ``closed'' if the set of spatial frequencies involved in its construction are the vertex of a polygon.
The minimal configuration to measure one closure phase is to use three telescopes, as the addition of several triplets of telescopes can describe al the possible polygon the information is clearly redundant. 
The rank of  $\bH^\psi$, the matrix which connects the phase closures to the polychromatic phases is given by the number of lines which ensure independent phase closures.
For a single analyzed wavelength the rank of the matrix $\bH^{\lambda_n}$ of section \ref{sec:T3Phi} is $N_\psi = {(N_t-1)(N_t-2)}\slash{2}$ \cite{meimon2005}.
When polychromatic data are measured the augmented matrix follows the right term of Eq. \ref{eq:clotphases}, the phase closures for each wavelength are estimated from the same set of telescopes triplets. 
Consequently, the rank of $\bH^{\bpsi}$ is then increased up by a factor $N_\lambda$: $\text{rank}(\bH^{\bpsi}) = N_\psi N_\lambda $. 
The ratio $\text{rank}(\bH^{\bpsi}) \slash  N_\phi = 1 - 2 \slash N_t  $ depends only on the number of telescopes.

For the differential phases, the construction of non redundant measures is detailed in section \ref{sec:DiffPhi} and is driven by $\bH^{\uDphi}$. 
The rank of this matrix is given by the number of independent differential phases that may result in the transformation from phases and is $\text{rank}( \bH^{\uDphi} ) = N_\bB (N_\lambda -1)$.
The ratio $\text{rank}(\bH^{\uDphi}) \slash N_\phi  =  1 - 1 \slash N_\lambda  $ depends only on the number of analyzed wavelengths.

As $\bH$ is the concatenation of $\bH^{\bpsi}$ and $\bH^{\uDphi}$ its rank is given by the number of independent relation between these matrices. 
Consider three telescopes $(a,b,c)$ and two wavelengths $(\lambda_1, \lambda_k)$, $\lambda_1$ is chosed to be the differential phases reference channel,
a linear relation which connects the phase closures of Eq. \ref{eq:hphi} to the differential phases of Eq. \ref{eq:hdphi} is:
\begin{equation}
\psi^{\lambda_k}_{a,b,c} - \psi^{\lambda_1}_{a,b,c} = 
\bdphi^{\lambda_k, \lambda_1}_{{a,b}}  + \bdphi^{\lambda_k, \lambda_1}_{{b,c}}  - \bdphi^{\lambda_k, \lambda_1}_{{a,c}}  \label{eq:T3DphiRelation} 
\end{equation}

the difference of lines of $\bH^{\bpsi}$ for a fixed wavelength only defines redundant phase closures.
The combination over $\lambda_k$  and $\lambda_1$ of all the triplets of bases, as in Eq. \ref{eq:T3DphiRelation}, expresses all the differential phases and so the rank of $\bH$ is at least the rank of $\bH^{\uDphi}$.
A more general phase closures differences $\psi^{\lambda_k}_{a,b,c} - \psi^{\lambda_m}_{a,b,c}$ can be expressed in term of $(\lambda_k,\lambda_1)$, leads to redundant information and adds nothing. 
Consequently, the only remaining independent information is monochromatic and corresponds to all the phase closures related to the reference channel $\lambda_1$
so $\text{rank}( \bH) = N_\bB (N_\lambda -1)  + N_\psi $.
Finally, the ratio $ \text{rank}(\bH)  \slash N_\phi = 1 - 2\slash( N_t N_\lambda) $ depends on the number of telescopes and analyzed wavelengths.

\begin{figure}[ht!] \vspace{-2mm}
 \includegraphics[width=\columnwidth]{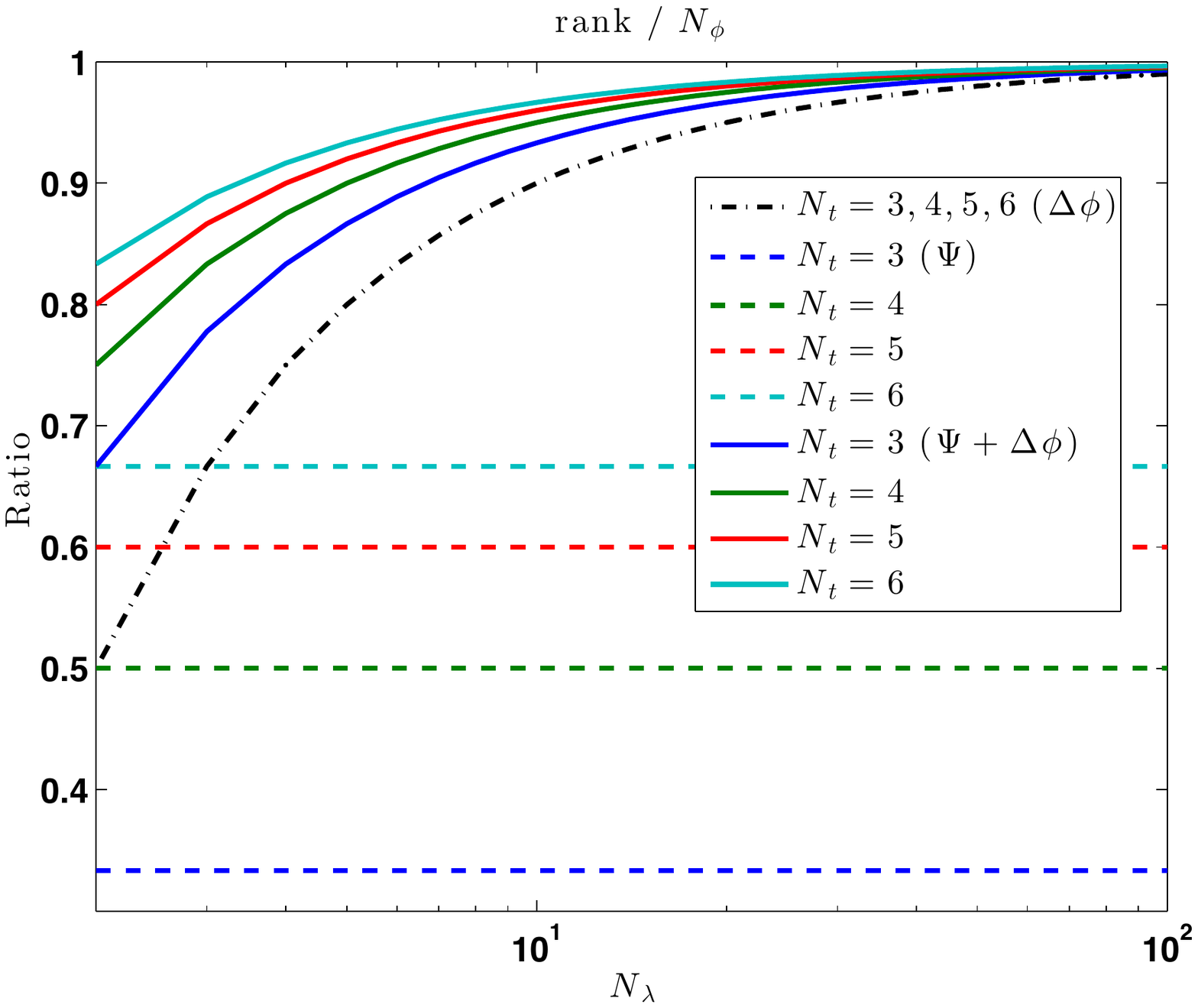} \vspace{-3mm}
 \caption{Ratio of the number of independent phases difference to the total number of phases as a function of the number of telescopes and analyzed wavelengths.
 }
 \label{Fig:Rank}  
 \end{figure}

The ratio of the number of independent phases difference for the two models (ranks of $\bH^{\bpsi}$ and $\bH^{\uDphi}$) and the augmented model (rank of $\bH$) to the number of phases is shown in Fig .\ref{Fig:Rank}. The ratio is drawn for $N_t=3,\ldots,6$ telescopes which corresponds to actual interferometers and up to $100~N_\lambda$. Note as an example that $N_\lambda > 500$ for the high resolution mode of AMBER \cite{AMBER}, and $N_\lambda =3$ for the standard uses of PIONIER \cite{PIONIER2011}.

 The use of all phases differences information largely increases the rank of the matrix which is always greater than for the phase closures alone.

\section{Inverse problem approach} \label{sec:InversePb}
\subsection{Image reconstruction problem formulation}
According to Eq.~\ref{eq:phasecomp} and notations defined in 
Eqs. \ref{eq:pow} and \ref{eq:phi}, a model for measurements of difference phases and squared moduli   can be written as 
\beq
 \uxi &=& \bH ~\uphi + \unoise_\xi \label{eq:measphase}\\
 \datapow &=& \upow^2 + \unoise_\zeta \label{eq:measmod}
\eeq
where $\unoise_\xi$ and $\unoise_\zeta$ account for noise and modeling errors.
Classical assumptions on their distributions are considered here.
The noise $\unoise_\zeta$ is assumed to be jointly independent and Gaussian \cite{meimon2005,LITPRO2008}, and
the noise $\unoise_\xi$ is assumed to be jointly independent and marginally  Von Mises distributed \cite{MIRA}.

Writing the opposite logarithm of the likelihood related to the squared absolute value and phase differences model, we obtain:
\beq
g^\text{data}(\ux) = \alpha \,  g^\zeta(\uy^\gamma) + \beta \, g^\xi(\uy^\phi) \label{eq:cout}
\eeq
where, $\alpha$ and $\beta$ are relative weighting terms and: 
\beq
&&g^\zeta(\uy^\gamma) = 
\sum_n \frac{1}{\uw_n} \lbr \datapow_n - \upow_n^2 \rbr^2
 \label{eq:Jmod}
, \, \upow = | \uuF \ux | \\
&&g^\xi(\uy^\phi) =  - \sum_m \boldsymbol{\kappa}_m \cos{\lbr \bh_m \uphi - \uxi_m \rbr}   \label{eq:Jphase}
, \, \uphi = \angle\;  \uuF \ux 
\eeq
where $\uw_n$ is the variance of $\datapow_n$. The constants  $\boldsymbol{\kappa}_m$ are related to the variance of $\uxi_m$
by $$\text{var}(\uxi_m)=1-I_1(\boldsymbol{\kappa}_m)/I_0(\boldsymbol{\kappa}_m)$$where $I_j$ is the modified Bessel function of order $j$ \cite{mardia}.
In practice $\boldsymbol{\kappa}_n$ is computed inverting numericaly the previous equation.
The variance of the closure phase and differential phases, if not provided by the instrument pipeline,
are estimated assuming independence of phase measurements.

Image reconstruction can be seen as an inverse problem \cite{IImgercons,InversePBTarantola05,InversePBThiebaut}.
The model which connects object to the measured data involves 
a NuDFT, as described in Eq.~\ref{eq:TensorEq} and Eqs.~\ref{eq:measphase}  and \ref{eq:measmod}. 
This transformation leads to a poor coverage of the spatial frequency plan and makes the problem ill-conditioned which requires tackling the image reconstruction as a regularized optimization problem \cite{RenardReg}. We will adopt here an objective function  of the form: 
\beq
\ux \leftarrow \underset{\ux \in  \support }{\text{minimize}} \lbr g^\text{data}(\ux) +  f^\text{reg}(\ux) \rbr 
\label{eq:gdatafreg}
\eeq
where the image $\ux$ can be constrained to be spatially limited in the support $\support$, and further constraints such as non negativity can be added in
$f^\text{reg}(\ux)$, which contains all the regularization terms. The support constraint is not included in $f^\text{reg}(\ux)$
for technical reasons related to the ADMM methodology described below.

\subsection{Regularizations and constraints}

OI images are by nature non negative  and sometimes contain sources that are spatially localized.
However, specifying the properties of the object parameters $\ux$ only in terms of non negativity and   spatial support is usually not  constraining enough.  It follows that the uses of regularization terms to emphasize some inherent a priori knowledge about the image structure is necessary. 

Following the matrix notation  for the 3D object as defined in Eq.~\ref{eq:ModMatX}, 
\logo\  in its current form  accounts the following classical priors:
\begin{itemize}
\item Ridge/Tikhonov regularization, motivated by the poor conditioning of 
the NuDFT operator.
\item Spatial or spectral Total Variation \cite{babacan09}.
\item Spatial or spectral smoothness  \cite{admmthiebaut}.
\end{itemize}
The support constraint is 
defined by the parameters space $\support$ in Eq.~\ref{eq:gdatafreg}
and the non-negativity constraint by the regularization term $\indic_{\mathbb{R}^+}(\uX)$.
In this description the regularization function in Eq.~\ref{eq:gdatafreg} writes
\beq
f^\text{reg}(\ux) &=&
\indic_{\mathbb{R}^+}(\uX) 		\,+\,
\lbr \mu_\varepsilon  \slash 2 \rbr \| \uX \|_\text{F}^2 	\,+\, \cdots\\ &&   \mutv \Omega_\ptv( \bHtv \uX)	  \,+\,
\muss \Omega_\pss( \uX\bHss)	 \nonumber
  \label{eq:RegAll}
\eeq
$\bHtv$ and $\bHss$ are the matrices of finite difference, associated with the spatial and spectral regularizations respectively
\cite{admmthiebaut}. 
$\bHtv \uX$ act on the column of $\uX$ which are images processed independently while $\uX\bHss$ operates on  the line of $\uX$ to connects the pixels between wavelengths.
 $\Omega_\ptv(\cdot)$ and $\Omega_\pss(\cdot)$ are matrix regularization terms that can be chosen as:
\beq
  \label{eq:regterms}
  \Reg_{1}(\bM)=\sum_{i,j}|\bM_{i,j}|\qquad
  \Reg_{2}(\bM)=\sum_{i,j}\bM_{i,j}^2 
\eeq
Finally, $\mu_\varepsilon$, $\mutv$ and $\muss$ are hyper-parameters which control the weights of the associated regularization terms.

\section{3D reconstruction algorithm} \label{sec:3Dimages}
Owing to the unavoidable non convexity of the problem as defined by Eq.~\ref{eq:gdatafreg} (see e.g.  in \cite{Meimon:05}),
the vast majority of image reconstruction algorithms  rely on a descent  optimization principle. So does \logo\, by using the flexibility of the Alternate Direction Methods of Multipliers (ADMM) algorithm, which
was  already used in \cite{admmthiebaut,admmsoulez}  to reconstruct  stellar spectrum of point 
sources from complex visibilities.  
Within this framework the  reconstruction algorithm will iterate as follow:
\begin{enumerate}
\item Update the complex visibilities.
\item Use the estimated complex visibilities and spatio--spectral regularization to reconstruct the polychromatic object (3D--images).
\item Update the Lagrange multipliers.
\end{enumerate}
Specific details are given in the rest of this section.
\subsection{ADMM optimization algorithm} \label{sec:admm}
The ADMM framework allows to split complex problem into smaller and easier ones by introducing auxiliary variables. However, in this case additional terms have to be taken into account. 
The subproblems can be solved independently by means of proximal operators. 
Standard constraints are taken into account in \logo\, and most proximal operators used in this section are known.
The optimization problem of Eq.~\ref{eq:gdatafreg} where 
$g^\text{data}(\ux)$ is given by Eqs.~\ref{eq:cout}--\ref{eq:Jphase}
and the regularization term $f^\text{reg}(\ux)$ is given by  Eq.~\ref{eq:RegAll}
is equivalent to:
\beq
\begin{aligned}
& \underset{\ux\in\support,\ux_i,\uy,\uy^\gamma,\uy^\phi}{\text{minimize}}
& &  \alpha \ g^\zeta(\uy^\gamma) + \beta \ g^\xi(\uy^\phi) + \frac{\mu_{\varepsilon}}{2}  \| \uX \|_\text{F}^2 +\indic_{\mathbb{R}^+}(\uX)
\\ & &&+\mutv\Reg_\ptv( \bHtv \uX) \nonumber
+\muss\Reg_\pss( \uX\bHss)
\label{eq:admm} \\
& \text{subject to}
& & \uy^\gamma =\uy,\uy^\phi=\uy,\uX_i=\uX, \uy= \uuF ~\ux 
\end{aligned}
\\ 
\eeq
In this equation,  the vector $\uy^\gamma$ denotes the complex visibilities variables whose likelihood is associated with the measurements of
visibilities squared absolute values, Eq.~\ref{eq:Jmod}.
The vector $\uy^\phi$ denotes the complex visibilities variables whose 
likelihood is associated with the measurements of
the visibilities phases {\it differences}, Eq.~\ref{eq:Jphase}.

Using the same approach for the three last regularization terms leads to the introduction of new auxiliary variables
$\uP$, $\uT$ and $\uV$ and to replace each regularization term in Eq.~\ref{eq:admm} by a constrained problem.
\begin{itemize}
\item The non-negativity constraint $\indic_{\mathbb{R}^+}(\uX)$ and the support constraint in Eq.~\ref{eq:admm}  are replaced by
\beq
\begin{aligned}
\underset{\uP}{\text{minimize}} \quad & \indic_{\mathbb{R}^+}(\uP)\\
\text{subject to}\quad & \uP = \uX, \uP\in\support
\end{aligned} \label{subind}
\eeq
\item The total variation regularization $\Reg_\ptv( \bHtv \uX)$ in Eq.~\ref{eq:admm}  is replaced by
\beq
\begin{aligned}
\underset{\uT}{\text{minimize}} \quad & \Reg_\ptv(\uT)\\
\text{subject to}\quad & \uT = \bHtv \uX 
\end{aligned} \label{subTV}
\eeq 
\item The spectral regularization $\Reg_\pss( \uX\bHss)$ in Eq.~\ref{eq:admm}  is replaced by
\beq
\begin{aligned}
\underset{\uV,\uS}{\text{minimize}} \quad & \Reg_\pss( \uV)\\
\text{subject to} \quad & \uV =  \uS \bHss \,\text{and}\,\, \uS = \uX  
\end{aligned} \label{subSS}
\eeq
the use of two auxiliary variables ensure spectral separability when solving for $\uX$.
\end{itemize}

The final optimization problem obtained by replacing Eqs. \ref{subind}, \ref{subTV} and \ref{subSS} in Eq.~\ref{eq:admm} is iteratively solved
using the ADMM algorithm \cite{ADMMBoyd}. 
Auxiliary variables related to the complex visibility: $\uy$, $\uy^\gamma$, $\uy^\phi$ have a proper Lagrange multiplier:
$\utau_y$ (or $\uTau_y$, with columns ${\uTau}^{\lambda_n} _y$,  in matrix form), $\utau_\gamma$, $\utau_\phi$ and share the same augmented Lagrangian parameter $\rho_y$.
The auxiliary variables introduced by the regularization: $\uP$, $\uT$, $\uV$, $\uS$ are associated with the Lagrange multipliers
$\uTau_P$, $\uTau_T$, $\uTau_V$, $\uTau_S$ and to the augmented Lagrangian parameters $\rho_P$, $\rho_T$ and $\rho_S$ for $\uV$ and $\uS$.

Denoting with a ``+'' superscript updated quantities, alternated minimization of the augmented Lagrangian gives:
\begin{align}
&\uy^{\gamma\,+} = \arg \min_{\uy^\gamma}  \ \alpha \ g^\zeta(\uy^\gamma)+ 
\frac{\rho_y}{2}  \|  \uy^\gamma - \uyt^\gamma \|^2_2  \label{updategam} \\
&\uy^{\phi\,+} = \arg \min_{\uy^\phi}  \ \beta \ g^\xi(\uy^\phi)+ 
\frac{\rho_y}{2}  \|  \uy^\phi - \uyt^\phi\|^2_2 \label{updatephi} \\
&\uy^+ = \frac{1}{3}\lbr \uy^{\gamma \ + } + \uy^{\phi \ +} + \uuF\ux + \rho_y^{-1} 
(\utau_\uy - \utau_\phi - \utau_\gamma)\rbr \\
&\uX^{\lambda_n \ +} = 
{\uC^{\lambda_n}}^{-1}   \left[ {\uF^{\lambda_n} }^H \lbr \rho_y {\uY}^{\lambda_n \ +}  - 
{\uTau}^{\lambda_n} _y
\rbr +  \cdots \right. 
\nonumber\\
& \qquad {\bH^\TotV}^\top  \lbr \rho_\tT \uT^{\lambda_n}  - \uTau^{\lambda_n} _\tT  \rbr +  \cdots \nonumber\\ &  \qquad  \left.  \lbr \rho_\tP\,
{\uP^{\lambda_n}} - \uTau^{\lambda_n} _\tP  \rbr   +  
\lbr \rho_\tS \, {\uS^{\lambda_n}} - \uTau^{\lambda_n} _\tS  \rbr   \right] \label{updateX}\\
&\ux^+=\text{vec}\, \uX^+\\
&\uP^+ = \arg \min_{\uP\in\support} \indic_{\mathbb{R}^+}(\uP) + \frac{\rho_\tP}{2} \left \|  \uP -  \widetilde{\uP} \right \|^2_2 \label{updateps} \\
&\uT^+ =\arg \min_{\uT} \mutv\Reg_\ptv( \uT) + \frac{\rho_\tT}{2} \left \|  \uT -\widetilde{\uT}\right \|^2_2 \label{updatetv} \\
&\uS ^+  =  \lb {\bHss} {\bHss}^\top  + \Id_{N_\lambda}\rb^{-1 } \lb \lbr \uV -\uTau_\tV / \rho_\tS  \rbr {\bHss}^\top + \right.
 \cdots \nonumber\\ & \left. \qquad \lbr \uX^+ +\uTau_\tS / \rho_\tS \rbr \rb\\
&\uV^+ = \arg \min_{\uV}  \muss\Reg_\pss( \uV) + \frac{\rho_\tS}{2} \left\| \uV -\widetilde{\uV}\right\|^2_2  \label{updatess} 
\end{align}
with the definitions:
\begin{align}
& \uyt^\gamma = \uy+ \rho_y^{-1}\utau_\gamma \\
& \uyt^\phi = \uy+\rho_y^{-1}\utau_\phi  \\
& \uC^{\lambda_n} = \rho_y {\uF^{\lambda_n} }^H \uF^{\lambda_n} + \rho_\tT  {\bHtv}^\top \bHtv +  \cdots \nonumber\\ & 
\lbr \mu_\varepsilon + \rho_\tP + \rho_\tS\rbr \Id_{N^2_x}\\
& \widetilde{\uP}  = \uX^+ + {\uTau_\tP} \slash {\rho_\tP}  \\
&\widetilde{\uT} = \bHtv \uX^+ + {\uTau_\tT} \slash {\rho_\tT} \label{Ttilde}\\
&\widetilde{\uV}=  \uS^+ \bHss  +\uTau_\tV \slash \rho_\tS \label{Vtilde}
\end{align}
The  update of the Lagrange multipliers are: 
\begin{align}
&\utau^+_{\gamma} = \utau_\gamma+ \rho_y   \lbr  \uy^+ - \uy^{\gamma \ +}  \rbr \\
&\utau^+_{\phi} =  \utau_\phi+ \rho_y   \lbr  \uy^+ - \uy^{\phi \ +}  \rbr \\
&\utau^+_{y}= \utau_y+ \rho_y   \lbr    \uuF \ux^+  - \uy^+ \rbr \\
& \uTau^+_\tP = \uTau_\tP + \rho_\tP \lbr  \uX^+  - \uP^+ \rbr \\
& \uTau^+_\tT = \uTau_\tT + \rho_\tT \lbr  \bHtv \uX^+ -  \uT^+  \rbr \\
& \uTau^+_\tS = \uTau_\tS + \rho_\tS  \lbr  \uX^+ - \uS^+ \rbr  \\
&\uTau^+_\tV = \uTau_\tV + \rho_\tS \lbr  \uS^+ \bHss -  \uV^+  \rbr 
\end{align}

The proximal operators \cite{CombettesPesquet} in Eqs.~\ref{updategam}  and \ref{updatephi} are detailed in subsections 
\ref{sec:sectionV2} and \ref{sec:sectionXi} respectively and the proximal operators for the regularization terms   in
 Eqs.~\ref{updateps}, \ref{updatetv} and \ref{updatess} are detailed in subsections \ref{sec:proxpos}--\ref{sec:proxtv}.

\subsection{Proximal operator for squared visibility} \label{sec:sectionV2}
The proximal operator of Eq.~\ref{updategam} updates the estimation of the complex visibilities from the measured squared absolute visibilities. 
Replacing $g^\zeta(\uy^\gamma)$ by its expression in Eq.~\ref{eq:Jmod}, Eq.~\ref{updategam} separates on each component:
\begin{equation}
{\uy^\gamma}^+_n =\arg \min_{\uy^{\gamma}_n}  ~ 
\alpha \frac{1}{\uw_n} {\lbr \datapow_n - \upow_n^2 \rbr^2}+  \frac{\rho_y}{2} | \uy_n^{\gamma} - \uyt_n^{\gamma} |^2  \label{eq:proxV2}
\end{equation}
which separates again on the modulus and phase of $\uy^{\gamma}$ as:  
\beq \begin{aligned}
{\upow}^+_n &=  \arg \min_{\upow_n>0} ~ \frac{\alpha}{\uw_n} \lbr \datapow_n - \upow_n^2 \rbr^2 +  \frac{\rho_y}{2} \lbr \upow_n - \widetilde{\upow}_n \rbr^2  \\
{\uphi}^+_n  &=  \widetilde{\uphi}_n  
\end{aligned} \eeq
where $\widetilde{\upow}$ and $\widetilde{\uphi}$ denote in this section the moduli and phases of $ \uyt^\gamma$.

The minimization of the previous fourth order polynomial for $\upow_n>0$ is obtained by computing the real roots of its derivative: 
\begin{align}
& \upow_n^3 + \lbr ~\rho_y \frac{ \uw_n }{4 \, \alpha}   -  \datapow_n  \rbr~\upow_n - \rho_y \frac{ \uw_n }{4 \, \alpha} \, \widetilde{\upow}_n  = 0 
\label{eq:cubicsystem}  
\end{align} 
using Cardano's method \cite{Jacobson09}. $\upow_n^+$ is the real positive root of 
Eq.~\ref{eq:cubicsystem} 
which minimizes the criterium Eq.~\ref{eq:proxV2}.  If Eq.~\ref{eq:cubicsystem} has no positive roots, then the polynomial to minimize is strictly 
increasing for $\upow_n>0$ and $\upow_n^+=0$. 

\subsection{Proximal operator for phase differences} \label{sec:sectionXi}
The proximal operator of Eq.~\ref{updatephi} updates the estimation of the complex visibilities from the measured visibilities phase differences. 
This problem involves triplets and duets of visibility phases which, contrarily to the moduli, are not separable.
Replacing  $g^\xi(\uy^\phi)$ given in Eq.~\ref{eq:Jphase} we obtain:
\begin{align}
{\uy^{\phi}}^+ = & \arg \min_{\uy^\phi} ~ 
 -\beta~  \sum_m \kappa_m \cos{\lbr \bh_m \uphi - \uxi_m \rbr} 
 +  \cdots \nonumber\\ & 
 \frac{\rho_y}{2} \sum_n \lbr \upow_n^2 + \widetilde{\upow}_n^2 -  2 \ \upow_n{\widetilde{\upow}}_n \cos{\lbr \widetilde{\uphi}_n - \uphi_n \rbr} \rbr
 \label{updateyphi}
\end{align} 
where $\widetilde{\upow}$ and $\widetilde{\uphi}$ denote in this section the modulus and phases of $ \uyt^\phi$.
Minimization of Eq.~\ref{updateyphi} w.r.t. $\upow_n>0$ gives:
\begin{align}
\upow_n^+(\uphi_n)= \max \lbr 0, {\widetilde{\upow}}_n \cos{\lbr \widetilde{\uphi}_n - \uphi_n \rbr} \rbr \label{updategammaphi}
\end{align} 
Eq.~\ref{updategammaphi} is first replaced in Eq.~\ref{updateyphi} and the resulting function minimized w.r.t. $\uphi$.
This minimization is carried out numerically using  a gradient descent algorithm. 
A compact expression of the gradient of the function to
be minimized is:
\begin{align}
\beta \, \bH^T \text{diag}(\boldsymbol{\kappa})^{-1} \sin{\lbr \bH \uphi - \uxi \rbr} - \rho_y~  \uPow^+(\uphi) {\widetilde{\uPow}}  \sin{\lbr \widetilde{\uphi} - \uphi \rbr} 
\end{align} 
where $\uPow^+(\uphi)=\text{diag}(\upow^+(\uphi))$ and $\widetilde{\uPow}=\text{diag}(\widetilde{\upow})$ are diagonal matrices with elements $\upow^+(\uphi)$ and $\widetilde{\upow}$. 
Once $\uphi^+$ is obtained, $\upow^+$ is given by $\upow^+(\uphi^+)$.
It is worthy to note that if at an iteration step of the descent algorithm $|\widetilde{\uphi}_n - \uphi_n|>\pi/2$,
from Eq.~\ref{updategammaphi} the corresponding term in the second part of Eq.~\ref{updateyphi} reduces to $\widetilde{\upow}_n^2$ and the 
next estimate of $\uphi$ will not depend on $\widetilde{\upow}_n$ and $\widetilde{\uphi}_n$.
For this reason the initial estimate of $\uphi$ is set at $\widetilde{\uphi}$. 
The minimization in \logo\, relies on the quasi-Newton algorithm implemented in \cite{minfunc}, note that the non convexity of the problem leads to a local minimization to estimate $\uphi$.

\subsection{Positivity and compactness} \label{sec:proxpos}

The solution of the constrained minimization in Eq.~\ref{updateps} \cite{ADMMBoyd, CombettesPesquet} is:
\begin{align}
& \uP^+_{k,l} = \left\{
    \begin{array}{ll}
        \widetilde{\uP} _{k,l}& \mbox{if } \widetilde{\uP} _{k,l} \odot \support_{k,l} >0 \\
        0 & \mbox{else.}
    \end{array}
\right. 
\end{align}
where $\support$ is a binary mask matrix (composed of $0$ and $1$) which represents the support of the object (disk etc.), each column of $\support$ is for a wavelength.
If the object support is not constrained  $\support_{k,l} = 1$ $\forall \, \lb{k,l}\rb$.

\subsection{Spatial and spectral regularization} \label{sec:proxtv}

The proximal operator of Eqs. \ref{updatetv} and \ref{updatess} is: 
\begin{itemize}
\item the \textit{soft thresholding} \cite{ADMMBoyd} if $\ptvss = 1$:
\beq
\begin{aligned}
&\uD^+_{k,l} = \left\{
    \begin{array}{ll}
        \uU_{k,l}& \mbox{if } \uU_{k,l} >0 \\
        0 & \mbox{else.}
    \end{array}
\right.\\
&\text{where } \uU_{k,l} = \lbr 1 - \frac{\mutvss}{|\widetilde{\uD}_{k,l}|} \rbr \widetilde{\uD}_{k,l}  
\label{proxtv}
\end{aligned}
\eeq
\item if $\ptvss_. =  2$: 
\begin{align}
\uD^+ = \frac{1}{1+2~\mutvss} \widetilde{\uD} 
\end{align}
\end{itemize}
where $\uD$ can be $\uT$ or $\uV$. $\ptv$ and $\pss$ can be chosen according to a priori selected on image, if the desired spatial/spectral smoothness regularization must promotes brightness jump then  $\ptvss =1 $ otherwise $\ptvss =2 $.
As an example, the proximal operator for the $\ell_2$ norm spectral smoothness regularization, as expressed in Eq.~\ref{updatess} with $\pss =  2$, is:
\begin{equation}
\uV^+ = \frac{1}{1+2~\muss} \widetilde{\uV} 
\end{equation}

Note that according to the structure of matrix $\widetilde\uT$, see Eq.~\ref{Ttilde}, the spatial proximal operator Eq.~\ref{proxtv} will apply
separately on images at different wavelengths. Similarly, according to the structure of matrix $\widetilde\uV$,
the spectral proximal operators operates separately on the voxels. 
Finally, in order to scale properly the regularization parameters, the $\mutv$ 
and $\muss$ of Eq.~\ref{eq:admm} are divided by their number of non zero elements in $\bHtv$ and $\bHss$ respectively.
This normalization should makes the parameters independent of the size of the 3D--image.

\section{Simulations} \label{sec:simus}
Computer simulations are presented in order to illustrate the performances of \logo\ and the benefits of a 
combined use of phase closures and differentials phases.  
To do so, synthetic data set are used and the algorithm conforms to the complete chain needed in OI image reconstruction i.e going from files to 3D--image.

\subsection{Synthetic data}
\subsubsection{Improvement related to phases differences}

\begin{figure}
\includegraphics[width=.61\columnwidth]{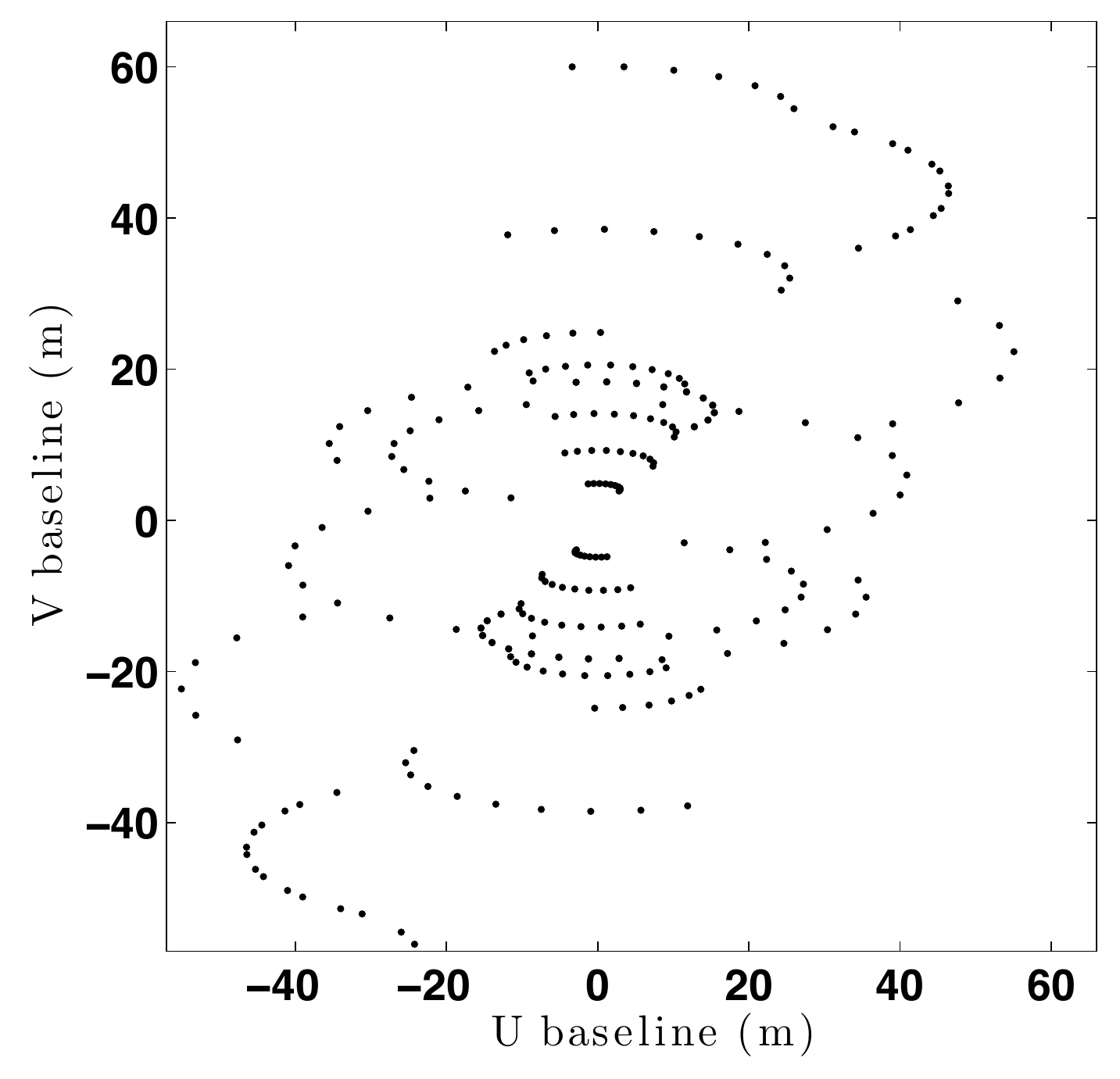} 
\vspace{-4mm}
\caption{Spatial frequencies coverage plan. Geometrical configuration  the 2004 International Beauty Contest in Optical Interferometry. 
}
\label{Fig:PlanUV1}  
\includegraphics[width=.9\columnwidth]{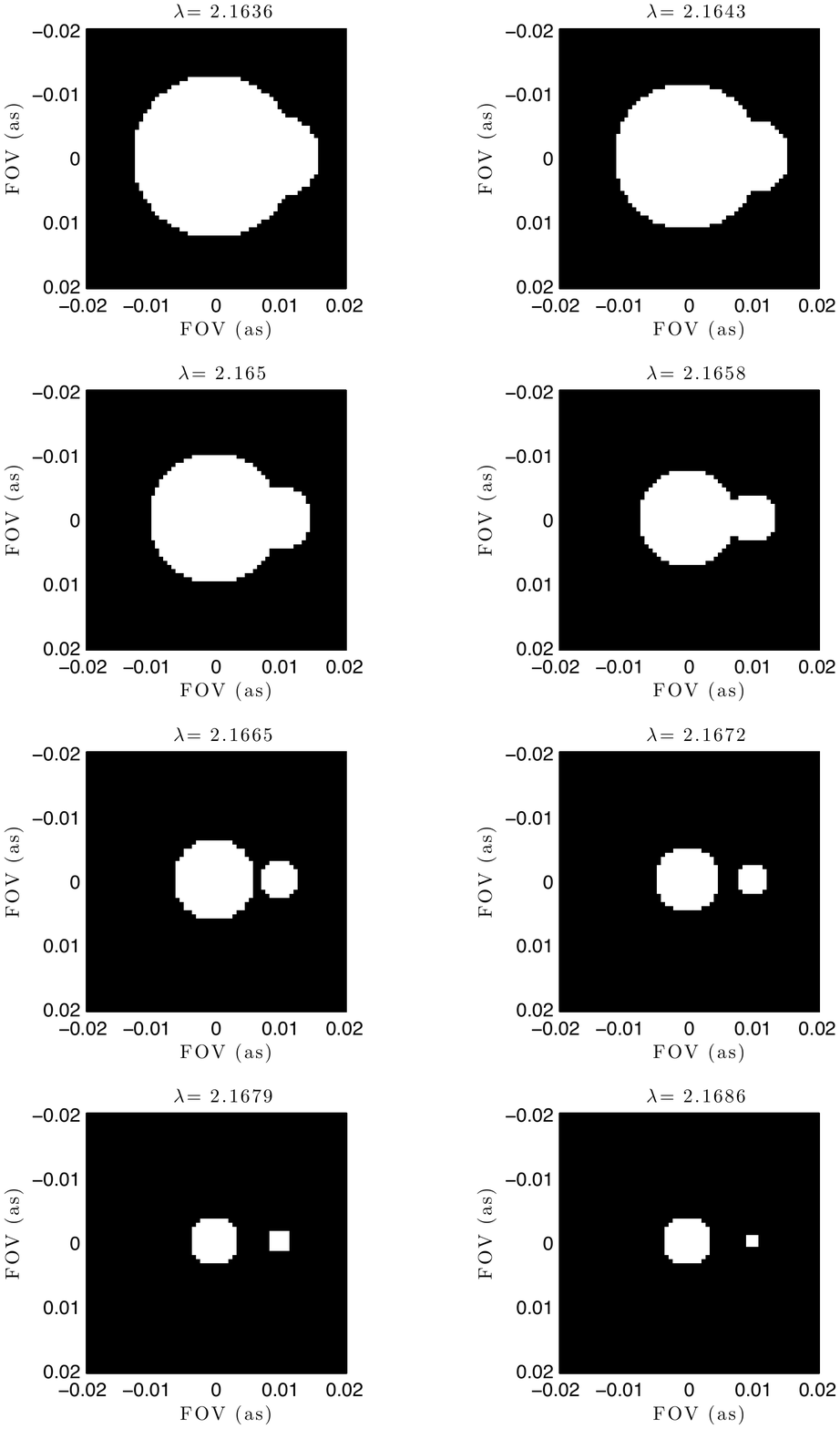} 
\vspace{-5mm}
\caption{Synthetic data, two disks with a decreasing radius along wavelength. Per channel view.}
\label{Fig:SolSynth}
\vspace{-9mm}
\begin{minipage}[l]{.49\columnwidth}
\includegraphics[width=\textwidth]{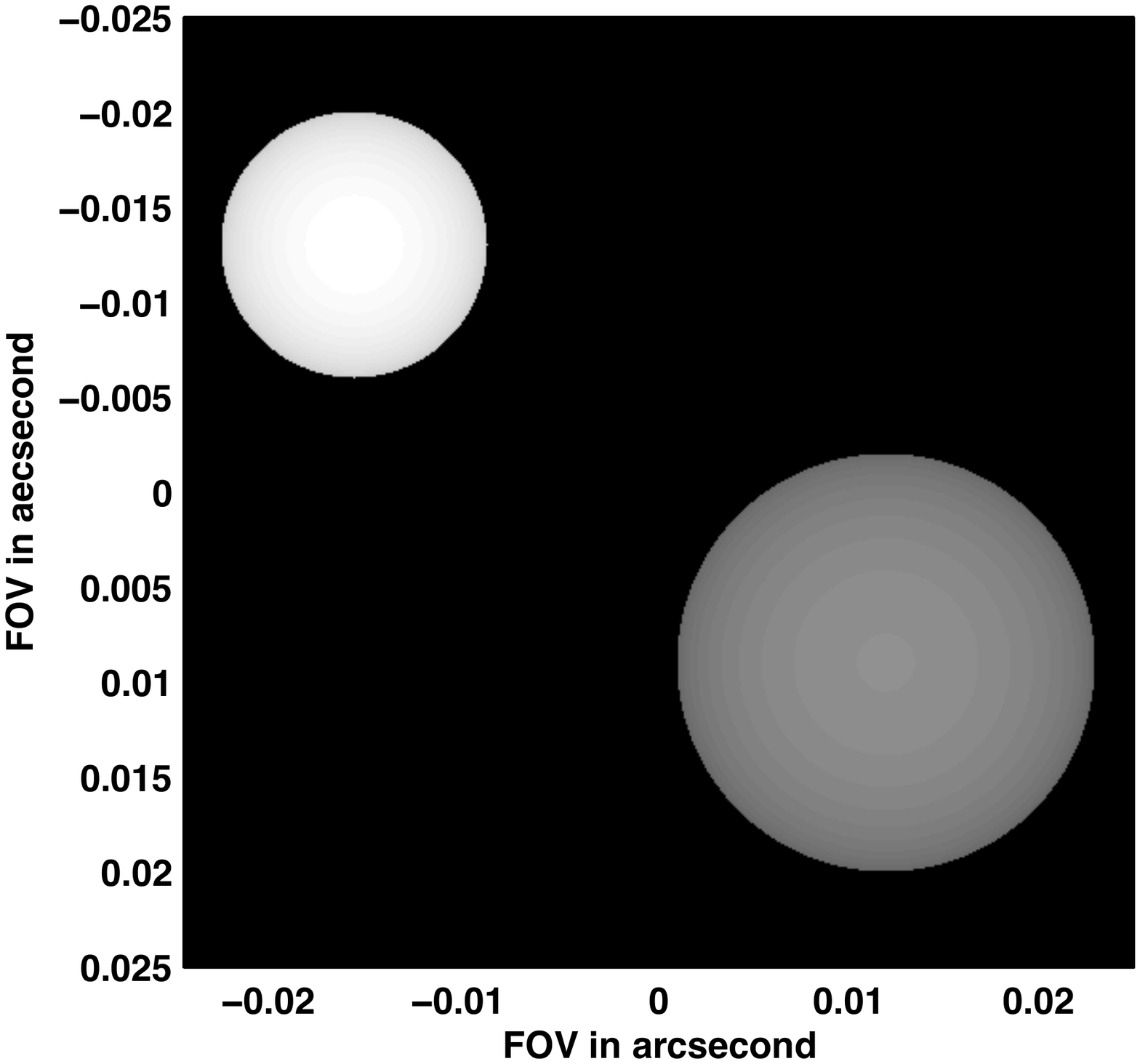} 
\end{minipage}
\begin{minipage}[l]{.45\columnwidth}
\includegraphics[width=\textwidth]{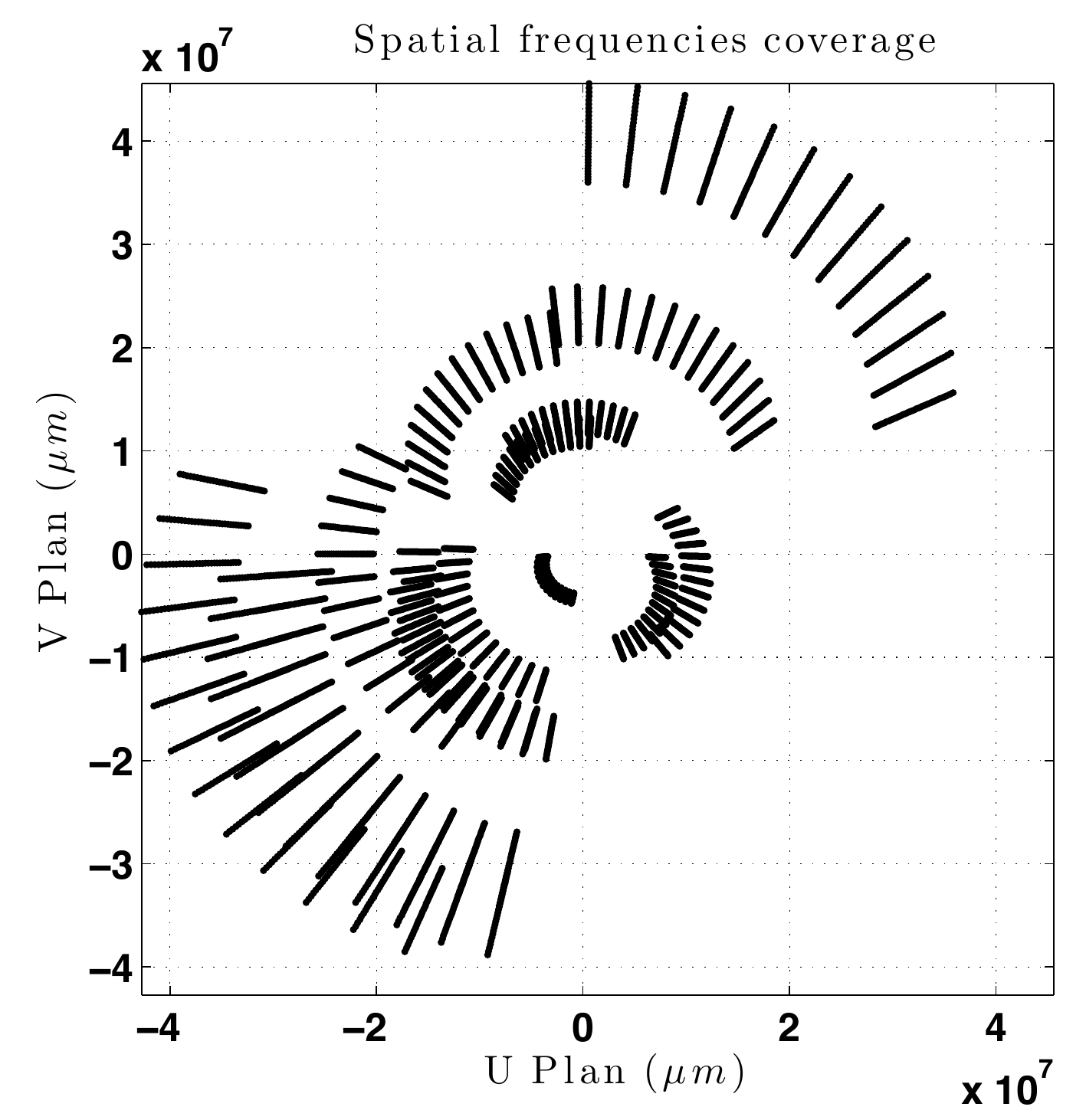} 
\end{minipage}  \vspace{-1cm}
\caption{Left: A channel of the 3D image used in simulation.
Right: Spatial frequencies coverage plan. Geometrical configuration  the 2004 International Beauty Contest in Optical Interferometry. 
}
\label{Fig:Obj2}
\label{Fig:PlanUV}  
\end{figure}

The first  simulation aims at determining the improvement on the reconstruction due of the combination of phases differences model. 
The simulated noiseless object consists in two uniform disks with a decreasing radius with wavelengths. The intensity of each disk as a function of the wavelength is constant. 
The original cube is of size $64 \times 64$ pixels $\times 8$ channels. 
 The instrumental configurations (geometrical position, wavelengths, acquisition time,...) 
 have been generated with the ASPRO2 \cite{ASPRO} software which simulates realistic interferometric 
 observation and store the data into OIFITS files \cite{OIFITS}.
 The simulation uses the configuration of the AMBER instrument at the VLTI \cite{AMBER} with three telescopes.
 Height equispaced wavelengths in the range $2.1635 \ \mu  m - 2.1686 \ \mu m$ (high resolution) at five acquisition instants are analyzed.
 The spatial frequencies coverage, including the earth rotation effect, is shown in Fig. \ref{Fig:PlanUV1} 
 for the spectral channel at $\lambda= 2.16 \ \mu  m$.
 This results in the measurement of 129 squared absolute visibilities and 43 phase closures for each wavelength.  
 The differential phases are calculated as in Eq.~\ref{eq:phaseDiffRef} and corresponds to $128\times (8-1)=903$ measures. 
 The object is shown in Fig. \ref{Fig:SolSynth} in a channel per channel view.

\subsubsection{Resolved stars reconstruction}

The second simulation deals with realistic noisy synthetic data with a given signal to noise ratio, SNR = 30 dB. 
More specifically, to generate the data we add zero-mean independent Gaussian random noise to each measurement. 
For each phase data $\varphi$, the standard deviation of the noise is $\sigma_\varphi$ = 1 rd/SNR (in radians). 
For each amplitude data $\gamma$, the standard deviation of the noise is $\sigma_\gamma$ = $\gamma \slash$SNR.
The simulation consists in two resolved stars of different spectral types: G8V and M3V for the large and small stellar disks respectively.  
The spectra and their chromatic limb darkening laws are based on Kurucz and van~Hamme models \cite{Kurucz,VH,refId0}. 
A channel of the original 3D-image is shown in Fig. \ref{Fig:Obj2}, . 
The two disks have a diameters of $22$ and $14$ milli-arcsecond (mas) and individual brightness distribution is varying among wavelengths. 
The original cube with a pixel resolution of $0.1$ mas is of size $501\times501$ pixels $\times30$ channels, the field of view (FOV) is $\approx 50$ mas.
The instrumental configuration was used in the 2004 International Beauty Contest in Optical Interferometry \cite{IBC04}: 
Thirty equi-spaced wavelengths in the range $1.45 \ \micron - 1.84 \ \micron$ (high resolution) at 13 acquisition instants are analyzed.
 The spatial frequencies coverage, including the earth rotation effect, is shown in Fig. \ref{Fig:PlanUV}.
 This results in the measurement of 195 squared absolute visibilities and 130 phase closures for each wavelength.  
 The differential phases are calculated as in Eq.~\ref{eq:phaseDiffRef} and corresponds to $130\times (30-1)=3770$ measures and simulation data are stored into OIFITS files \cite{OIFITS}.

\subsection{Reconstruction parameters}

The initial estimate was a Dirac function centered on the image for each wavelength and the algorithm was stopped after $5000$ iterations.

 \begin{figure}[ht!]
 \includegraphics[width=.79\columnwidth]{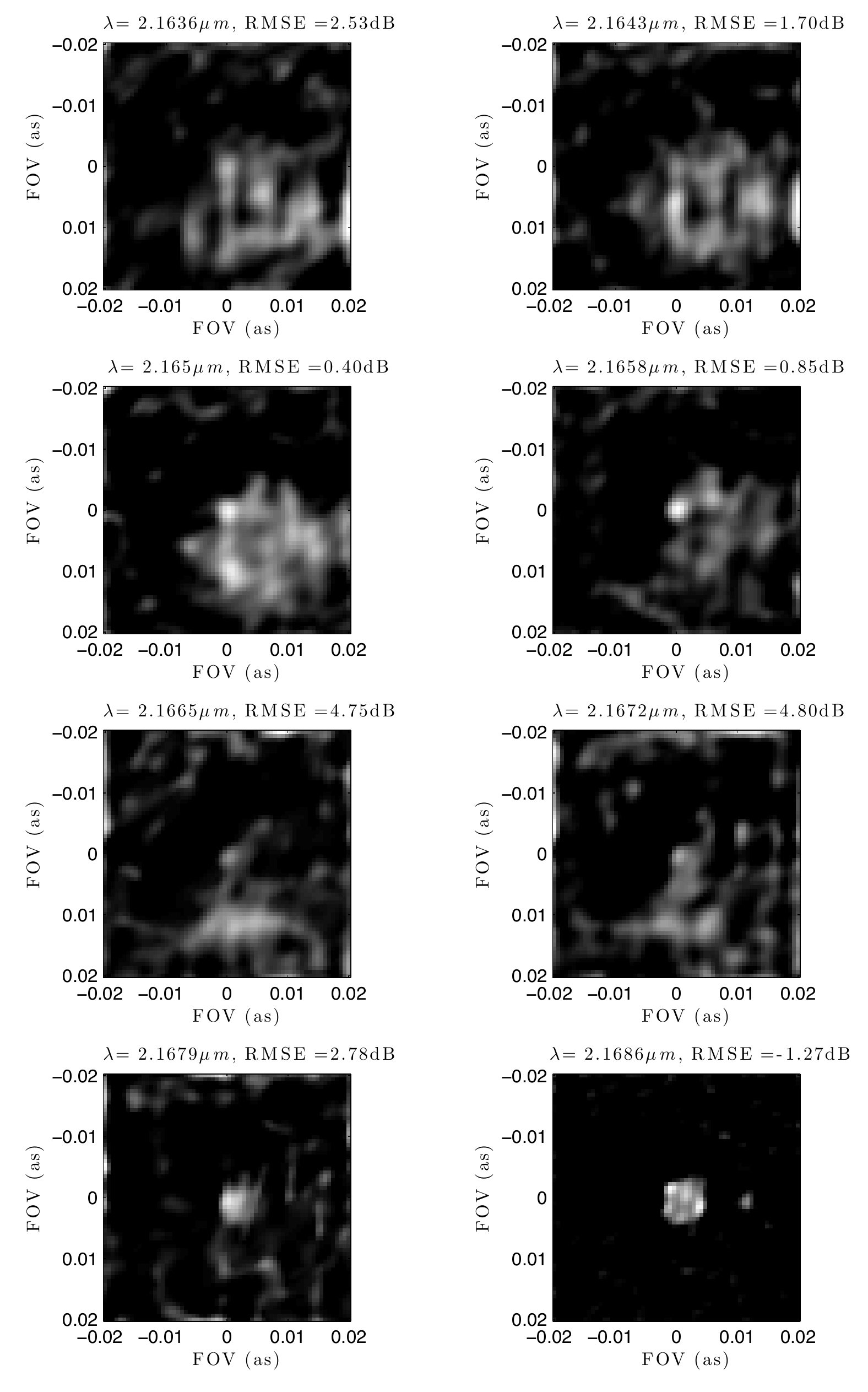}
   \vspace{-4mm}
 \caption{Estimated 3--D object with closure phases only.}
 \label{Fig:EstSynthT3}  
 \end{figure}

The type of regularization was chosen according to the a priori on the object: $\ptv = 1$ and $\pss = 2$  to enforce spatial continuity with sharp edges and spectral smoothness.  The regularization weights $\mutv = 2$ and $\muss= 2$, were empirically tuned to have a visually acceptable solution (though not the best one). The associated augmented parameters,  $\rho_\tT = 10$ and $\rho_\tS = 10$, which drive the  convergence of the spatial/spectral regularization sub-problems have been tuned to be not too small (to avoid divergence) and not too large (to not slow down the convergence). Without loss of generality the problem can be dimensionless in $\rho_y$ and thus we took $\rho_y=1$. We tuned the weights of the two terms of Eqs. \ref{updategam}  and \ref{updatephi} so that, at convergences, these terms are of the same order.  This yields  $\alpha= 10^2$ and $\beta= 10^2$. Finally, we took $\mu_\varepsilon  = 10^{-6}$.
The reconstruct cube is of size $64\times64$ pixels per channel ($N_\lambda=8$ and $N_\lambda=30$ for the first and the second simulation respectively), leading to $4096 \times N_\lambda$ parameters to estimate.

\subsection{Simulation results}
\subsubsection{Improvement related to phases differences}

In order to compare the improvement due to the combinations of phase differences we compare the results obtained using the squared visibility with: 
the closure phases, the differential phases and both phase differences.

 \begin{figure}[ht!]
  \includegraphics[width=.79\columnwidth]{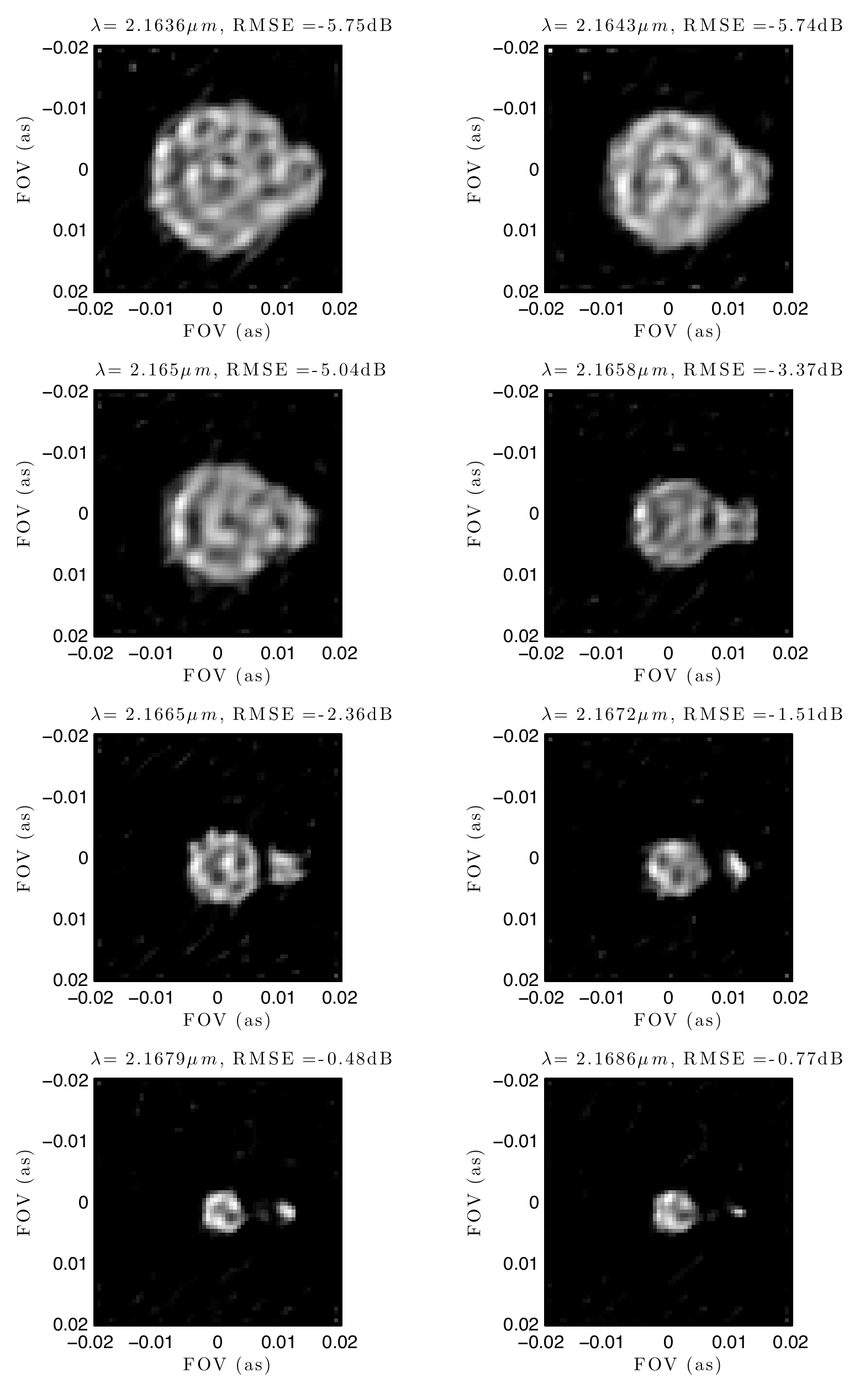}
  \vspace{-4mm}
\caption{Estimated 3--D object with differentials phases only.}
 \label{Fig:EstSynthDP}  
 \end{figure}

 The classical observable of phase differences used in OI images reconstructions is the closure phases.
 The first estimation uses only this phase information. The result is shown in Fig. \ref{Fig:EstSynthT3}. 
 Due to the lack of information: $43$ phase closures for $129$ baselines per wavelength, the reconstruction performances are quite poor.
 The narrow objects at high wavelength are well reconstructed. The object structure (two disks) is not correctly found at lower wavelengths where artifacts are present.

 The second simulation uses the differentials phases as the only phase differences measure.
 It leads to the estimated objects shown in Fig. \ref{Fig:EstSynthDP}. 
 The object structures are better reconstructed. However some artifacts are present and the object surface are not as smooth as they should be.

\begin{figure}[ht!]
  \includegraphics[width=.8\columnwidth]{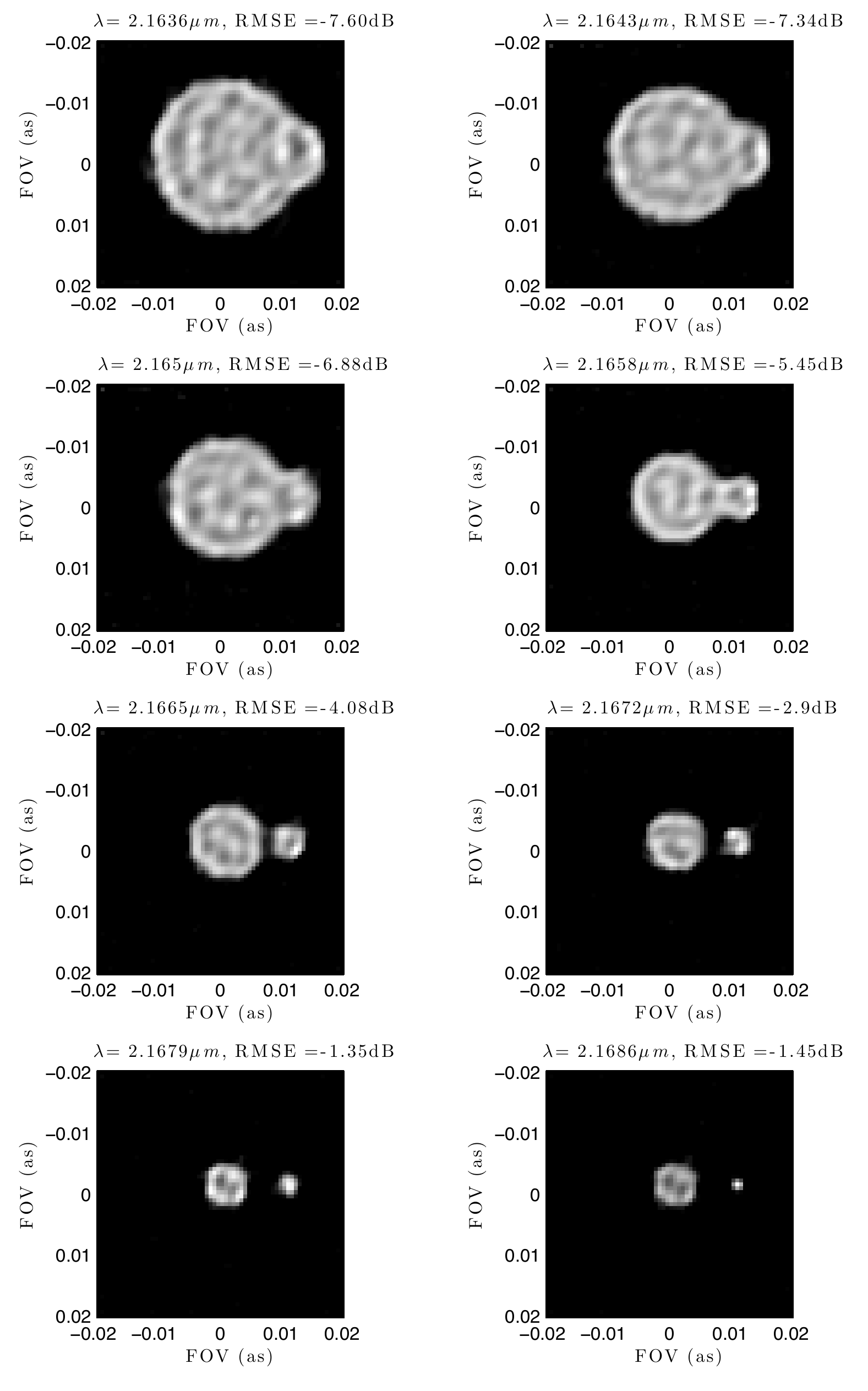} 
  \vspace{-1mm}
 \caption{Synthetic data. Estimated 3--D object with closure phases and differentials phases. }
 \label{Fig:EstSynthDPT3}  
 \end{figure}

 Finally, the last simulation uses all the available phase information: phase closures and differentials phases.
 The result in Fig. \ref{Fig:EstSynthDPT3} shows that the object is extremely well reconstructed at all wavelengths. 
The Relative Mean Squared Error (RMSE) of the reconstructed object ($\hat{x}$) at wavelength $\lambda_n$ defined as 
$\|  \textbf{x}^{\lambda_n} - \hat{\textbf{x}}^{\lambda_n}  \|_2^2 \slash \| \textbf{x}^{\lambda_n} \|_2^2$ has been computed for each wavelength.
The RMSEs in dB are presented in Fig. \ref{Fig:EstSynthT3}, \ref{Fig:EstSynthDP} and \ref{Fig:EstSynthDPT3} to quantify the improvement.
 
\subsubsection{Resolved stars reconstruction}
\begin{figure}[ht!]
\vspace{-20mm}
\includegraphics[ width=\columnwidth]{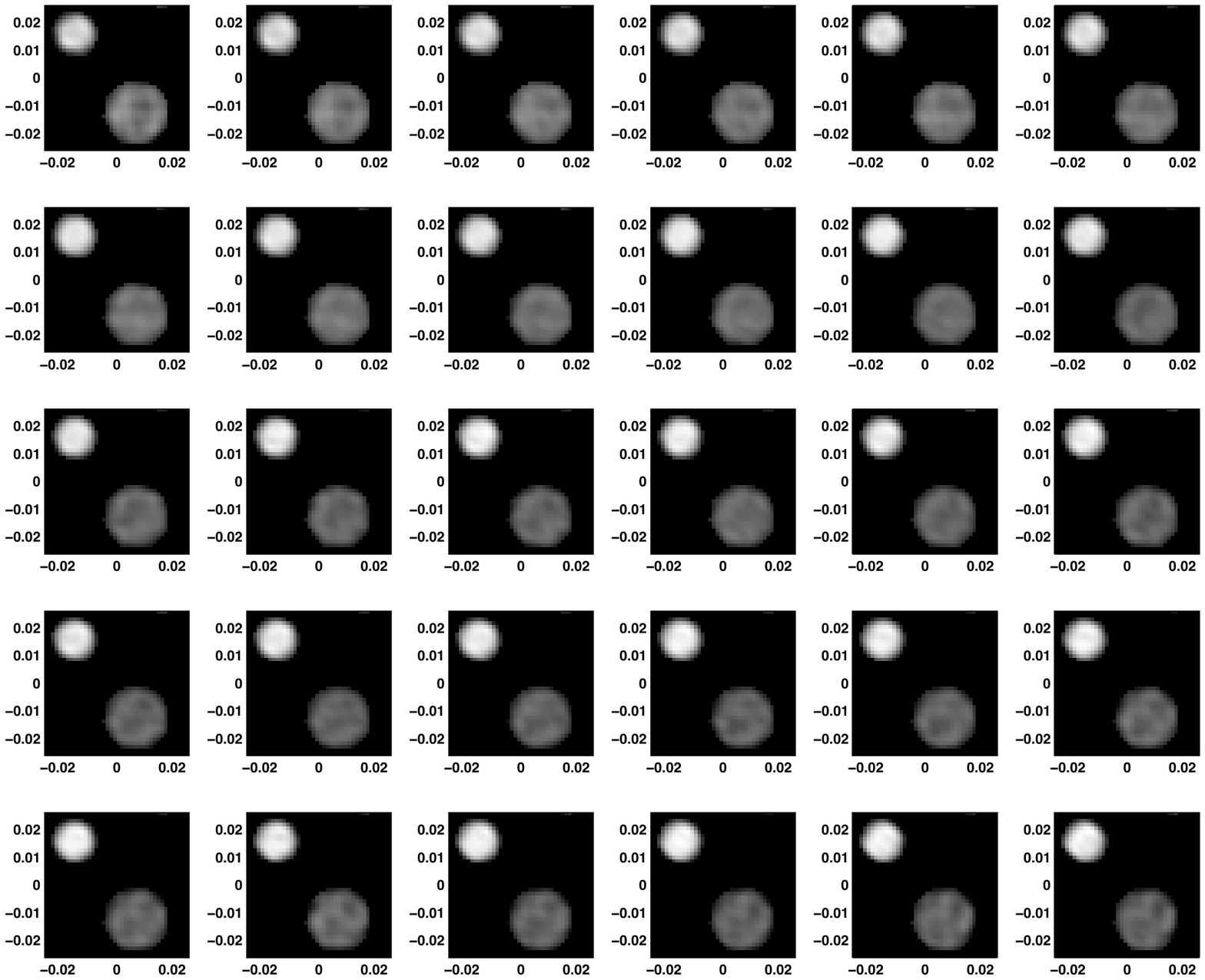}  \vspace{-30mm}
\caption{Estimated Images per channels, the units are in arc second.}  \vspace{1mm}
\label{Fig:Est3D}  
\includegraphics[width=\columnwidth]{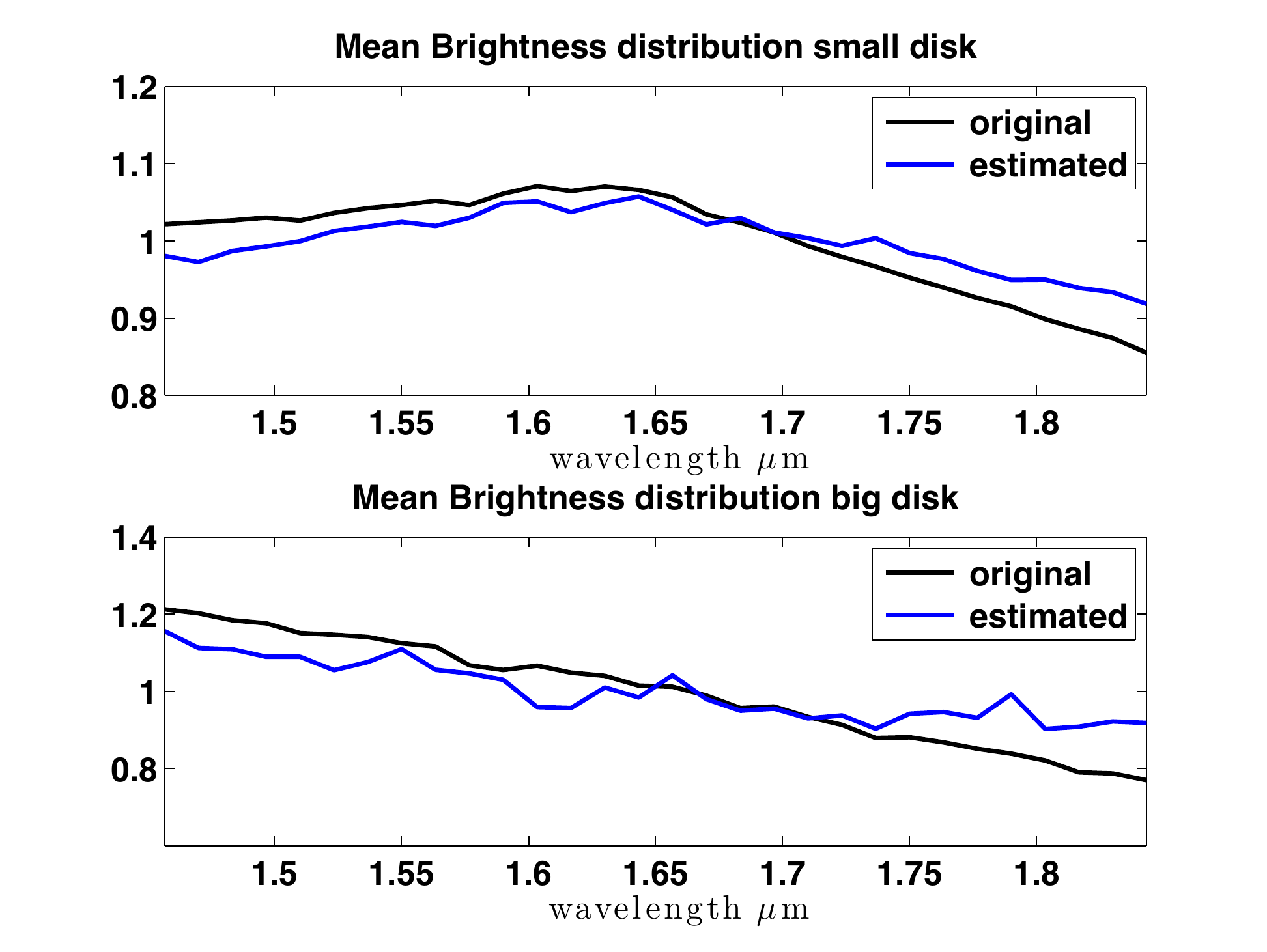}  
\caption{Original and estimated Brightness distribution overs channels for the two disks.}
\label{Fig:EstBD}    
\end{figure}

Images estimated per channel are shown in Fig. \ref{Fig:Est3D}, values lower than $1\%$ of the maximum of the cube are thresholded.
The shape and the size of the two stars are correctly restored.  
The polychromatic brightness distributions of the two estimated disks are compared to the solution in Fig. \ref{Fig:EstBD}.  
Even if the estimated images are not as smooth as the original ones the estimated brightness follows correctly the {\it true distribution}.

\section{Acknowledgments}
The present work was funded by the french ANR project POLCA (ANR-2010-BLAN-0511-02). One aim of POLCA is to elaborate dedicated algorithms for model-fitting and image reconstruction using polychromatic interferometric observations. The authors thank R. Petrov for proposing the use of the differential phase as a key item of this image reconstruction process.

\section{Conclusions}
A polychromatic 3--D image reconstruction for optical interferometry namely \logo\, is presented. 
The main contribution of \logo\,  is to uses both phase closure and differential phases information to estimates the original phases of the observed scene.
Compared to the use of phases difference independently, the combination leads to use a transformation matrix with a ``number of wavelengths/telescopes'' dependent rank.
The constraint optimization follows the ADMM methodology with spatial and spectral regularization. 
\logo\,  is able to deal with the complete chain of astronomical standard data (OIFITS file to data).
Realistic simulation on noisy data are encouraging and shows the potential of \logo.


\end{document}